\documentclass{aa}  

\usepackage{graphicx}
\usepackage{txfonts}
\usepackage{epsfig}
\usepackage{longtable}
\usepackage{lscape}
\usepackage{amssymb} 
\usepackage{hyperref}
\usepackage{refcount}

\def \inte {{\em INTEGRAL}}
\def \rxte {{\em RXTE}}

\def \suzaku {{\em Suzaku}}
\def \sw {{\em Swift}}
\def \xmm {{\em XMM--Newton}}

\def \hcm {\hbox {\ifmmode $ atom cm$^{-2}\else atom cm$^{-2}$\fi}}

\def \arcsec {\hbox{$^{\prime\prime}$}}

\def \ATel {ATel}

\def \apj {ApJ}
\def \apjl {ApJL}
\def \apjs {ApJS}\def \aap {A\&A}

\def \mnras {MNRAS}
\def \ssr {Space Science Reviews}

\begin{document} 

\title{Soft X-ray characterisation of the long term properties of Supergiant Fast X--ray Transients\thanks{Tables
\ref{sfxt10:tab:alldata08408}--\ref{sfxt10:tab:alldata_outburst} are  
available in electronic form at the CDS via anonymous ftp to  {\tt cdsarc.u-strasbg.fr} (130.79.128.5) or via 
 \href{http://cdsweb.u-strasbg.fr/cgi-bin/qcat?J/A+A/XXX/XXX}{{\tt http://cdsweb.u-strasbg.fr/cgi-bin/qcat?J/A+A/XXX/XXX}    } 
             } 
}

\titlerunning{Soft X--ray long term properties of SFXTs} 
\authorrunning{Romano et al.}

\author{P.\ Romano\inst{1}
           \and
          L.\ Ducci\inst{2,3} 
         \and
          V. Mangano\inst{4}
         \and
         P. Esposito\inst{5}  
         \and 
         E.\ Bozzo\inst{3} 
         \and
         S. Vercellone\inst{1} 
          }
   \institute{INAF, Istituto di Astrofisica Spaziale e Fisica Cosmica - Palermo,
              Via U.\ La Malfa 153, I-90146 Palermo, Italy \\
              \email{romano@ifc.inaf.it}
                \and 
             Institut f\"ur Astronomie und Astrophysik, Eberhard Karls Universit\"at, 
             Sand 1, 72076 T\"ubingen, Germany
                \and 
             ISDC Data Center for Astrophysics, Universit\'e de Gen\`eve, 16 chemin d'\'Ecogia, 1290 Versoix, Switzerland
            \and
              Department of Astronomy and Astrophysics, Pennsylvania State 
              University, University Park, PA 16802, USA
               \and 
             INAF, Istituto di Astrofisica Spaziale e Fisica Cosmica - Milano, 
              Via E.\ Bassini 15,   I-20133 Milano,  Italy 
              }
\date{Received 21 March 2014; accepted 21 June 2014 }

\abstract{
Supergiant Fast X--ray Transients (SFXTs) are High Mass X--ray Binaries (HMXBs) 
characterised by a hard X--ray   ($\geq 15$\,keV) flaring behaviour. These flares 
reach peak luminosities of 10$^{36}$--10$^{37}$~erg~s$^{-1}$ and last  
a few hours in the hard X--rays.  
}{
We investigate the long term properties of SFXTs 
by examining the soft (0.3--10\,keV) X-ray emission of the three least active SFXTs in the hard X--ray
and by comparing them with the remainder of the SFXT sample. 
}{
We perform the first high-sensitivity soft X--ray long-term monitoring with \sw/XRT of three 
relatively unexplored SFXTs, IGR~J08408$-$4503, IGR~J16328$-$4726, and IGR~J16465$-$4507, 
whose hard X--ray duty cycles are the lowest measured among the SFXT sample. 
We assess how long each source spends in each flux state and compare their
properties with those of the prototypical SFXTs. 
}{
The behaviour of IGR~J08408$-$4503 and IGR~J16328$-$4726
resembles that of other SFXTs, and it is characterized by a relatively high inactivity 
duty cycle (IDC) and pronounced dynamic range (DR) in the X-ray luminosity. 
We found DR$\sim$7400, IDC$\sim$67\,\% for IGR~J08408$-$4503, 
and DR$\sim$750, IDC$\sim$61\,\% for IGR~J16328$-$4726  
(in all cases the IDC is referred to the limiting flux sensitivity of XRT, 
i.e.\ 1--3$\times$10$^{-12}$\,erg cm$^{-2}$ s$^{-1}$).
In common with all the most extreme SFXT prototypes 
(IGR~J17544$-$2619, XTE~J1739$-$302, and IGR~J16479$-$4514), 
 IGR~J08408$-$4503 shows two distinct populations of flares.
The first one is associated with the brightest outbursts (X--ray luminosity 
$L_{\rm X}$$\gtrsim$10$^{35-36}$~erg s$^{-1}$), while the second comprises 
less bright events with typical luminosities $L_{\rm X}$$\lesssim$10$^{35}$~erg s$^{-1}$. 
This double-peaked distribution of the flares as a function of the X--ray luminosity 
seems to be a ubiquitous feature of the extreme SFXTs. 
The lower DR of IGR~J16328$-$4726 suggests that this is an intermediate SFXT.
IGR~J16465$-$4507 is characterized by a low IDC$\sim$5\,\% and a relatively 
small DR$\sim$40, reminiscent of classical supergiant HMXBs. 
The duty cycles measured with XRT are found to be comparable with those reported 
previously by BAT and \inte, when the higher limiting sensitivities 
of these instruments are taken into account and sufficiently long observational 
campaigns are available. By making use of these new results and those we reported previously, 
we prove that no clear correlation exists between 
the duty cycles of the SFXTs and their orbital periods.
}{
The unique sensitivity and scheduling flexibility of \sw/XRT allowed us to carry out an efficient 
long-term monitoring of the SFXTs, following their activity across more than 4 orders of magnitude 
in X-ray luminosity. While it is not possible to exclude that particular distributions of the 
clump and wind parameters may produce double-peaked differential distributions in 
the X-ray luminosities of the SFXTs, the lack of a clear correlation between the duty cycles
and orbital periods of these sources make it difficult to interpret their peculiar variability 
by only using arguments related to the properties of supergiant star winds. 
Our findings favour the idea that a correct interpretation of the SFXT phenomenology 
requires a mechanism to strongly reduce the mass accretion rate onto the compact object 
during most of its orbit around the companion,
as proposed in a number of theoretical works. 
}


   \maketitle

\setcounter{table}{4}  
 \begin{table*}
 \tabcolsep 4pt   
 \begin{center}
 \caption{Summary of the {\it Swift}/XRT campaign on the three SFXTs in the new monitoring sample.\label{sfxt10:tab:campaign} }
 \begin{tabular}{llrrrrlll}
 \hline
 \hline
 \noalign{\smallskip}
Name &Nickname&Campaign &Campaign &N\tablefootmark{a} &XRT Net &Outburst &BAT \\  
     &  &     Start         &End             & & Exposure      & Dates   & Trigger   & \\
     &  &     (yyyy-mm-dd)  &(yyyy-mm-dd)    & & (ks)  & (yyyy-mm-dd) & Number  & \\
  \noalign{\smallskip}
 \hline
 \noalign{\smallskip}
IGR~J08408$-$4503  &J08408&  2011-10-20    & 2012-08-05          & 82 &  74.4  & -- &       --  \\  
IGR~J16328$-$4726  &J16328&  2011-10-20    & 2012-10-22          & 82 &  73.5  & 2011-12-29 &       510701  \\ 
                                  &           &  2013-09-02    & 2013-10-24          & 16 &  14.5  & -- &       --  \\ 
IGR~J16465$-$4507  &J16465&  2013-01-20    & 2013-09-01          & 65 &  58.6  & --  &  -- \\ 
  \noalign{\smallskip}
 \hline
 \noalign{\smallskip}
Total             &&             &                      &245                        &  221.0 &           &              & \\  
  \noalign{\smallskip}
  \hline
  \end{tabular}
  \end{center}
\tablefoot{
\tablefoottext{a}{Number of observations obtained during the monitoring campaign.}
}
  \end{table*}
%

	\section{Introduction\label{sfxt10:intro}}

Supergiant fast X--ray transients (SFXTs) are the most recently recognized 
\citep[e.g.][]{Sguera2005} class of High Mass X--ray Binaries (HMXBs). 
They are associated with OB supergiant stars via IR/optical spectroscopy, 
and display hard X--ray  ($\geq 15$\,keV) outbursts significantly shorter 
than those of typical Be/X--ray binaries,  characterised  
by bright flares (peak luminosities of 10$^{36}$--10$^{37}$~erg~s$^{-1}$)
lasting a few hours  \citep{Sguera2005,Negueruela2006}. 
These bright flares are often clustered together in longer outbursts, 
lasting from a few hours to a few days 
\citep[e.g.,][]{Romano2007,Rampy2009:suzaku17544,Romano2014:sfxts_catI}.   
Their outburst spectra in the hard X--rays resemble those of HMXBs 
hosting accreting neutron stars, with spectrally hard power laws 
combined with high energy cut-offs,  
therefore it is generally assumed that all SFXTs might host a neutron star, 
even if pulse periods have only been measured for a few SFXTs. 
Since their quiescent luminosity is of the order of $\sim 10^{32}$~erg~s$^{-1}$
\citep[e.g.][]{zand2005,Bozzo2010:quiesc1739n08408}, SFXTs display a quite characteristic 
dynamic range of 3--5 orders of magnitude. 
At the time of writing, the SFXT class consists of 14 sources 
\citep[e.g.][and references therein]{Romano2014:sfxts_catI} and about as many
candidates, that is, sources which have shown an SFXT-like flaring behaviour, 
but are still lacking a detailed classification of the optical companion. 
About 250 HMXBs are currently known to reside in our Galaxy and the Magellanic Clouds
\citep[][]{Liu2005:hmxb_LMC_SMC,Liu2006:hmxb_Gal,Krivonos2012:ibis_gal_survey},  
so the SFXT population is quickly becoming not only a peculiar, but also a relevant portion 
of the HMXB population. 

The detailed mechanisms responsible for the bright outbursts are still being debated. 
It is generally believed that they are related to either the properties of 
the wind from the supergiant companion 
\citep{zand2005,Walter2007,Negueruela2008,Sidoli2007} 
and/or the presence of mechanisms regulating or inhibiting accretion 
(\citealt[][]{Grebenev2007}, propeller effect; \citealt[][]{Bozzo2008}, magnetic gating). 
Recently, a model of quasi-spherical accretion onto neutron stars involving 
hot shells of accreted material above the magnetosphere 
\citep[][]{ElsnerLamb1977,Shakura2012:quasi_spherical,Shakura2013:off_states} has also been  
proposed.

The long-term behaviour of SFXTs -- away from the prominent bright outbursts -- is naturally best 
observed by monitoring instruments, such as the 
Imager on Board the INTEGRAL Satellite \citep[IBIS,][]{Ubertini2003} 
or the Burst Alert Telescope \citep[BAT,][]{Barthelmy2005:BAT} on board \sw{} 
\citep[][]{Gehrels2004} 
that have now gathered data spanning about a decade each.  
Due to their sensitivity limits, however, these monitors mostly catch only the 
very bright flares.  
The low fluxes characteristic of the states outside the bright outbursts could  only be 
studied extensively when a highly sensitive, soft X--ray  (0.2--10\,keV) telescope, 
the X--ray Telescope \citep[XRT, ][]{Burrows2005:XRT} on board \sw{} 
was used in a two-year long series of pointed observations,  
3--4 days apart, during a systematic study \citep[][]{Sidoli2008:sfxts_paperI} of 
IGR~J16479$-$4514, XTE~J1739--302, IGR~J17544$-$2619, and AX~J1841.0$-$0536 
(hereon J16479, J1739, J17544, and J1841, respectively).  
This first assessment of how long each source spends in each flux state
yielded unexpected results. Not only the time spent in outburst was a small 
fraction (3--5\,\%) of the total  \citep[][]{Romano2009:sfxts_paperV}, but also the four sources (which we shall
call {\it initial monitoring sample} hereon) were found to spend most of their time 
at mean fluxes two orders of magnitude below the bright flares, at 
luminosities in the $10^{33}$--$10^{34}$ erg s$^{-1}$ range. 
The sources were detected in the soft X--ray for the majority of 
pointings so that their duty cycle of {\it inactivity} 
\citep[][and references therein]{Romano2011:sfxts_paperVI} 
was relatively small (19--55\,\%), 
clearly at odds with with what is generally observed in the hard X--rays. 
These datasets also established the ubiquitous flaring activity at all intensities and 
all timescales probed  that were consistently observed with the XRT as well as during 
deep pointed observations with \suzaku\ \citep[e.g][]{Rampy2009:suzaku17544} and 
\xmm\ \citep[e.g.][]{Bozzo2010:quiesc1739n08408}. 

Further monitoring campaigns providing high-cadence, pointed observations 
for one or more orbital periods were performed 
on IGR~J18483$-$0311,  IGR~J16418$-$4532, and IGR~J17354$-$3255 
(hereon J18483, J16418, and J17354, respectively)  
to primarily study the effects of orbital parameters on the observed flare distributions
\citep[][]{Romano2010:sfxts_18483,Romano2012:sfxts_16418,Ducci2013:sfxts_17354}. 
We call this latter group {\it orbital monitoring sample}. 

In this paper we continue our in depth exploration of the long term 
properties of SFXTs\footnote{\href{http://www.ifc.inaf.it/sfxt/}{Project web page: http://www.ifc.inaf.it/sfxt/ .} } 
with three \sw/XRT monitoring campaigns 
providing the first year-long high-sensitivity soft  X--ray coverage 
of IGR~J08408$-$4503, IGR~J16328$-$4726, and IGR~J16465$-$4507. 
These three SFXTs, which we shall call the {\it new monitoring sample}, 
are probably the least studied among the SFXT population, 
hence the interest on each individual source, 
whose long term soft X--ray properties are presented here for the first time. 
These sources also show the 
lowest hard X--ray duty cycles \citep[][]{Ducci2010,Paizis2014}. 
In Sections~\ref{sfxt10:sample} and \ref{sfxt10:dataredu} we introduce our new 
monitoring sample, the observing strategy, and the analysis of both the new data 
and the archival ones on the initial and orbital monitoring samples. 
In Sect.~\ref{sfxt10:results} we exploit the long baseline to calculate the 
soft X--ray inactivity duty cycle and 
perform intensity-selected spectral analysis of the new monitoring sample.  
We also create the differential distributions of flux  and luminosity 
for the whole SFXT sample (10 sources) searching for clues on the underlaying emission 
mechanisms. 
In Sect.~\ref{sfxt10:discussion} we discuss our findings and 
in Sect.~\ref{sfxt10:conclusions} we summarise our results and draw our conclusions.

\setcounter{table}{5}  
\begin{table*} 
 \tabcolsep 2pt         
 \begin{center}
 \caption{Duty cycle of inactivity.}
 \label{sfxt10:tab:dutycycle}
 \begin{tabular}{rccccclccc}
 \hline
 \hline
 \noalign{\smallskip}
Name (Nickname) 
&  Limiting Rate\tablefootmark{a} & Limiting $F$\tablefootmark{a} & Limiting $L$\tablefootmark{a} &$\Delta T_{\Sigma}$  & $P_{\rm short}$ &  IDC 
                  & Rate$_{\Delta T_{\Sigma}}$   & Distance & Ref.  \\
               & (0.2--10\,keV)   &  (2--10\,keV) &  (2--10\,keV)  &  &  &  & (0.2--10\,keV)   &  & \\                  
                & ($10^{-3}$ c s$^{-1}$) &  ($10^{-12}$ erg cm$^{-2}$ s$^{-1}$)  &($10^{34}$ erg s$^{-1}$)
                &(ks) & (\%) &  (\%) &   ($10^{-3}$c s$^{-1}$) & (kpc)       &             \\  
  \noalign{\smallskip}
 \hline
 \noalign{\smallskip} 
{\it Initial Monitoring Sample} \\
IGR~J16479$-$4514 (J16479)  & 16  & 2.5   & 1.1    &  29.7 & 3   &  19   &  3.1$\pm$0.5    & 4.9 &  1  \\
XTE~J1739$-$302    (J1739)  & 13  & 1.6   & 0.18  &  71.5 & 10 &  39                             &  4.0$\pm$0.3    & 2.7 &  1  \\ 
IGR~J17544$-$2619 (J17544) & 12  & 1.4   & 0.21  &  69.3 & 10 &  55                             &  2.2$\pm$0.2    & 3.6 &   1   \\ 
AX~J1841.0$-$0536 (J1841) & 13  & 1.8   & 1.6    &  26.6  & 3  &  28                             &  2.4$\pm$0.4    & 7.8$\pm0.74$ &  2  \\ 
  \noalign{\smallskip}
{\it New Monitoring Sample} \\
IGR~J08408$-$4503  (J08408)        &     17   & 1.9         &  0.26 & 46.6    &7        &  67      & $7.2\pm0.6$   & 3.4$\pm$0.34 &  2       \\  
IGR~J16328$-$4726  (J16328)        &     14   & 2.7         &  2.5   & 47.5    &12      &  61      & $4.0\pm0.4$   & 6.5$\pm$3.5 &  3          \\  
IGR~J16465$-$4507  (J16465)        &     16   & 2.0         &  4.4   & 3.0      &0        &  5        & $14.6\pm0.4$ & 12.7$\pm$1.3 &  2\\  
  \noalign{\smallskip}
{\it Orbital Monitoring Sample} \\
IGR~J16418$-$4532  (J16418)        &      19  & 12.5   &  36     &  4.8     & 0  &  11   & $>9.2$\tablefootmark{b}  & 13 &   1 \\  
IGR~J17354$-$3255  (J17354)        &      14  & 2.2     &  3.3    &  7.8     & 1  &  33                             & $>4.6$\tablefootmark{b}  & 8.5   &   4 \\  
IGR~J18483$-$0311 (J18483)         &      11  & 1.8     & 0.24   &  11.8   & 0  &  27                             & $3.6\pm0.8$               & 2.83$\pm$0.05 &   5 \\  
  \noalign{\smallskip}
  \hline
  \end{tabular}
  \end{center}
\tablefoot{
Count rates (Col.~2) are in units of $10^{-3}$ counts s$^{-1}$ in the 0.2--10\,keV energy band.  
         Observed fluxes (Col.~3) are in units of $10^{-12}$ erg cm$^{-2}$ s$^{-1}$ 
         and luminosities (Col.~4)  in units of $10^{34}$ erg s$^{-1}$, both in the 2--10\,keV energy band. 
    $\Delta T_{\Sigma}$ (Col.~5) is the sum of the exposures accumulated in all observations, 
   each in excess of 900\,s, where only a 3-$\sigma$ upper limit was achieved;  
   $P_{\rm short}$  (Col.~6) is the percentage of time lost to short observations; 
   IDC  (Col.~7, detailed in Sect.~\ref{sfxt10:idc}) is the  {\it duty cycle of inactivity}, 
   the time each source spends  undetected down to a flux limit of reported in column 3; 
   Rate$_{\Delta T_{\Sigma}}$ (Col.~8, detailed in Sect.~\ref{sfxt10:spec_out_of_outburst})  
   is the observed count rate in the data for which no detections were obtained as single observations.
   Values for the initial monitoring sample were recalculated based on the whole length of the campaigns 
   \citep[][]{Romano2011:sfxts_paperVI}. 
\tablefoottext{a}{Based on a single 900\,s exposure.}
\tablefoottext{b}{3-$\sigma$ upper limit. }
}
\tablebib{
(1) \citet{Rahoui2008}; 
(2) \citet[][]{Coleiro2013:IR_IDs}; 
(3) \citet[][3--10\,kpc]{Fiocchi2012:16328-4726}; 
(4) \citet[][]{Tomsick2009:cxc17354}; 
(5) \citet{Torrejon2010:hmxbs}.
}
  \end{table*} 

	\section{Sample and Observations} \label{sfxt10:sample}
 
The monitoring campaign commenced on 2011 October 20 with a focus 
on \object{IGR~J08408$-$4503} and \object{IGR~J16328$-$4726}  
for one solar year, and continued in 2013 with one year on \object{IGR~J16465$-$4507}.  
Given our preliminary results at the end of 2012, we also 
collected further data on IGR~J16328$-$4726 during 2013 to improve the statistics. 

The transient IGR~J08408$-$4503 (hereon J08408) was discovered on 2006 May 15 during a 900\,s  bright flare that  
reached a peak flux of 250 mCrab  \citep[20--40\,keV, ][]{Gotz2006:08408-4503discovery}. 
It was sought in archival \inte\ observations \citep{Mereghetti:08408-4503}, 
which demonstrated a recurrent transient nature, with an earlier active state in 2003.  
The \sw/XRT refined position \citep{Kennea2006:08408-4503} led to an 
association with an O8.5Ib(f) supergiant star, HD~74194, 
\citep{Masetti2006:08408-4503} at a distance of $\sim 3$\,kpc. 
\sw\ caught several bright flares from this source 
\citep[][and references therein]{Romano2013:atel5190}.

 The transient  IGR~J16328$-$4726 \citep[][hereon J16328]{Bird2007:igr3cat} has a long history of hard X-ray  
activity characterised by  flares lasting up to a few hours \citep[][]{Fiocchi2010:16328-4726} 
as observed by \inte.
\sw\ also caught one bright flare \citep[][]{Romano2013:Cospar12},
when the source reached an unabsorbed 2--10\,keV flux of $\sim 4\times10^{-10}$ 
erg cm$^{-2}$ s$^{-1}$. 
The orbital period is $P_{\rm orb}=10.076\pm0.003$\,d \citep{Corbet2010:16328-4726}, 
and the IR/optical counterpart is 2MASS~J16323791$-$4723409 \citep{Grupe2009:16328-4726}, 
an O8Iafpe supergiant star \citep[][]{Coleiro2013:IR_IDs}.

The source IGR~J16465$-$4507 (hereon J16465) was discovered by \inte\ on 2004 September 6--7, when it averaged  
$8.8\pm0.9$\,mCrab (18--60\,keV) and subsequently showed a strong flare at $\sim 28$\,mCrab
on September 7. It never triggered the \sw/BAT.  
IGR~J16465$-$4507 is a pulsar with $P_{\rm spin}=228\pm6$\,s \citep{Lutovinov2005}
and orbital period $P_{\rm orb}=30.243\pm0.035$\,d  \citep{LaParola2010:16465-4507_period}.
The optical counterpart is 2MASS J16463526$-$4507045 \citep[][]{ZuritaHeras2004:16465-4507} 
a B0.5Ib \citep{Negueruela2007}  star at a distance of about 8\,kpc 
\citep[but also see][]{Nespoli2008,Rahoui2008}. 

For these sources we obtained 2 observations week$^{-1}$ object$^{-1}$, 
each 1\,ks long.  The XRT mode was set in AUTO for J08408 and J16328
to best exploit XRT automatic mode 
switching \citep{Hill04:xrtmodes} in response to changes in the observed fluxes,
and in photon counting  (PC) mode for J16465. 
The observation logs are reported in 
Tables\footnote{Online only.} \ref{sfxt10:tab:alldata08408}, \ref{sfxt10:tab:alldata16328}, and 
\ref{sfxt10:tab:alldata16465}. 
We also considered data obtained while  J08408, J16328, and 
J1841.0 were in outburst to include in our count rate distributions
(Table \footnote{Online only.} \ref{sfxt10:tab:alldata_outburst}).
During this new monitoring campaign
we collected a total of 245 \sw\ observations as part of our program, for a 
total net XRT exposure of $\sim 221$\,ks accumulated on the three sources and 
distributed as shown in Table~\ref{sfxt10:tab:campaign}.

        \subsection{Reanalysis of the initial and orbital monitoring samples\label{sfxt10:olddata}}

We considered the data on the initial monitoring sample, 
J16479, J1739, and J17544, collected during the first two years of monitoring 
\citep[2007-10-26 to 2009-11-03]{Romano2011:sfxts_paperVI}, 
and those on J1841 collected during the first year of monitoring  
\citep[2007-10-26 to 2008-11-15]{Romano2009:sfxts_paperV}. 
We also considered the data on the orbital monitoring sample, 
J18483 \citep[2009-06-11 to 2009-07-08]{Romano2010:sfxts_18483},
J16418 \citep[2011-02-18 to 2011-07-30]{Romano2012:sfxts_16418}, and  
J17354 \citep[2012-07-18 to 2012-07-28]{Ducci2013:sfxts_17354}. 

We reanalyzed them by using the most recent software and calibrations like the newly 
acquired data, as described below.

 	 \section{ Data reduction} \label{sfxt10:dataredu}

The XRT data were processed with standard procedures 
({\sc xrtpipeline} v0.12.6), filtering and screening criteria by using {\sc FTOOLS} (v6.13).  
During the monitoring campaigns the source count rates never exceeded 
$\sim0.5$ count s$^{-1}$, so only PC events (selected in grades 0--12) 
were considered. 
Source events were accumulated within a circular region 
with an outer radius of 20 pixels (1 pixel $\sim2.36$\arcsec).  
Background events were accumulated from an annular source-free region
centered on J08408   (inner/outer radii of 100/160 pixels), 
and on J16465  (inner/outer radii of 80/120 pixels), 
and from a nearby source-free circular region (80 pixels) for J16328. 
The data obtained during outbursts to include in the count rate distributions
were  affected by pile-up, and were corrected by adopting standard procedures
\citep[][]{vaughan2006:050315,Romano2006:060124}. 
The outburst data, reported in Table~\ref{sfxt10:tab:alldata_outburst}, 
come from the 2013 July 2 outburst for J08408 \citep[][]{Romano2013:atel5190},
the 2009 June 10 one for  J16328 \citep[][]{Romano2013:Cospar12}, 
and the 2012 June 14 one for J1841 \citep[][]{Romano2013:MI50x_sfxts}. 
For our spectral analysis, we extracted events in the same regions as 
those adopted for the light curve creation;  
ancillary response files were generated with {\sc xrtmkarf},
to account for different extraction regions, vignetting, and PSF corrections. 
We used the latest spectral redistribution matrices in CALDB (20130313). 
For a more detailed discussion of the data analysis procedure, we refer the reader to 
\citet[][and references therein]{Romano2011:sfxts_paperVI}.

The BAT data of the outburst of 2011 December 29   06:39:20 UT 
(image trigger number 510701)\footnote{See \citet[][]{Romano2013:Cospar12} for an analysis 
of the 2009 June 10 outburst.}  of J16328 
were analyzed using the standard BAT software  within {\sc FTOOLS}. 
The source is not detected above $\ga 70$\,keV. 
The BAT mask-weighted spectrum was extracted during the first orbit of data; 
an energy-dependent systematic error vector was applied and 
response matrices were generated with {\sc batdrmgen}.  
The spectrum was fit in the 15--70\,keV range with a simple power law, obtaining 
$\Gamma_{\rm BAT\,2011}=3.0\pm1.0$ ($\chi^2_{\nu}=1.162$,  37 d.o.f.). 
The 20--50\,keV flux was $2.8\times10^{-10}$ erg cm$^{-2}$ s$^{-1}$.

\begin{figure}
\begin{center}
\vspace{-2.0truecm}

\hspace{-1.truecm}
\centerline{\includegraphics[width=9.5cm,angle=0]{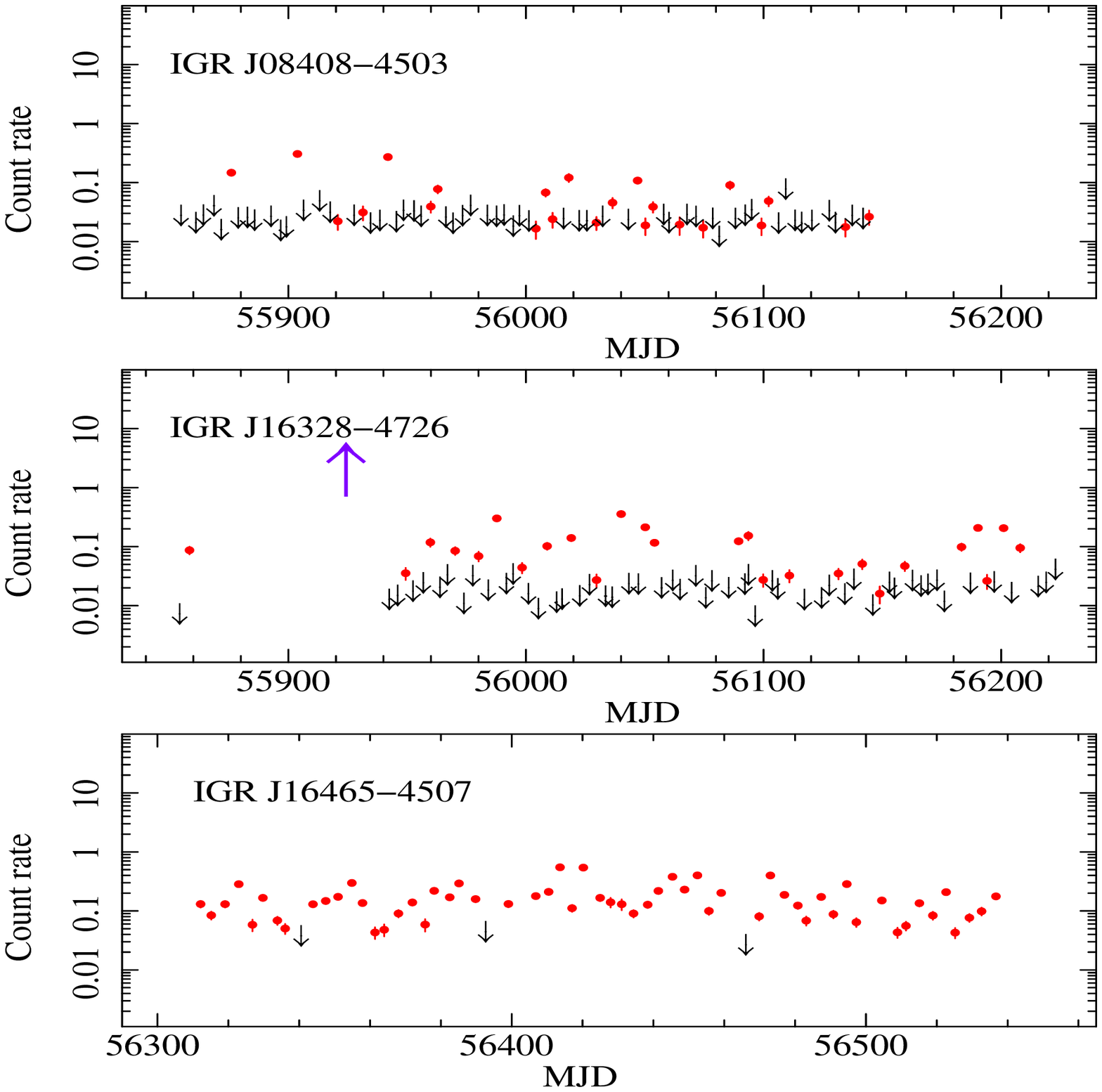}}
\vspace{-3.0truecm}
\caption[XRT light curves]{\sw/XRT (0.2--10\,keV) light curves.  
                The data were collected from 2011 October 20 to  2012 October 22
                and  from 2013 January 20 to October 24.  
		The (black) downward-pointing arrows are 3$\sigma$ upper limits. The upward pointing arrow 
                marks a flare that triggered the BAT on MJD 55924 (2011 December 29).  
 		\label{sfxt10:fig:xrtlcvs} 
}
\end{center}
\end{figure}
%
\section{Results} \label{sfxt10:results} 
        \subsection{Soft X--ray light curves and dynamical ranges\label{sfxt10:xrtlcvs}}

The 0.2--10\,keV XRT light curves  
are shown in Fig.~\ref{sfxt10:fig:xrtlcvs}. 
They are corrected for PSF losses and vignetting, and background-subtracted. 
Each point in the light curves refers to the average count rate observed
during each observation performed with XRT. Hereon errors on count rates are 
at the 1-$\sigma$ level.

The dynamic range (DR), which we shall define as the maximum to minimum ratio, 
in count rate units, is probably the simplest  piece of information we can measure 
from the X-ray light curves. Its knowledge has, however, quite a large impact 
in our understanding a source, since it can be used to discriminate 
 \citep[][]{Negueruela2006:ESASP604,Walter2006} between outbursts 
of classical supergiant HMXB (sgHMXB, $\lesssim 50$) 
and SFXT ($\gtrsim100$). 

For each source we calculated the observed XRT DR during this monitoring
when considering individual detections, 3$\sigma$ upper limits, 
and the peak count rate ever observed by XRT. 
For  J08408 we obtain a minimum DR of 25 
(the maximum value being $\sim 0.3$ counts s$^{-1}$, 
the minimum a 3$\sigma$ upper limit at 0.012 counts s$^{-1}$). 
The overall DR, considering the recorded outbursts 
\citep[][and references therein]{Romano2013:atel5190}      
reaches then about 2000. 
J16328 reaches a DR in excess of 50 
(the maximum value being $\sim 0.3$ counts s$^{-1}$, 
the minimum a 3$\sigma$ upper limit at 0.01 counts s$^{-1}$). 
The overall DR, considering the recorded outburst 
\citep[][maximum at $\sim 3$ counts s$^{-1}$]{Romano2013:Cospar12} 
then is of the order of $\sim 500$. 
J16465 was detected in all observations except 3, 
and shows a DR of 12, the peak being 0.55 counts s$^{-1}$. 
The DR only increases to 20 if individual 3$\sigma$ upper limits are considered
(the lower being 0.026 counts s$^{-1}$), as this source never triggered the BAT. 
By considering the detections obtained by combining all data for each source where 
individual observations only yielded 3$\sigma$ upper limits (see below Section~\ref{sfxt10:idc}
and Col.\ 8 in Table~\ref{sfxt10:tab:dutycycle}), the overall DR are $\sim 7400$,  
$\sim 750$,  and  38, 
for J08408, J16328, and J16465, respectively.

\setcounter{table}{6}  
\begin{table*}
 \tabcolsep 2pt         
 \begin{center}
 \caption{XRT spectroscopy of the three SFXTs in the new monitoring sample (2011--2013 data set).\label{sfxt10:tab:xrtspecfits} }
 \begin{tabular}{crrcccccccclccr}
  \hline
\hline
 \noalign{\smallskip}
Name     & Spectrum  & Mean Rate               &$N_{\rm H}$              & $\Gamma$  &   $kT_{\rm BB}$ &$R_{\rm BB}$  &Flux\tablefootmark{a}   & Luminosity\tablefootmark{b} &$\chi^{2}_{\nu}$/dof  & F-test\tablefootmark{c} & Fig.  \\ 
&           &  &   &   &   & &(2--10 keV)  & (2--10 keV)  &  &   $p$ & \\
&           & (c s$^{-1}$)& ($10^{22}$~cm$^{-2}$) &   & (eV) &(km) &($10^{-12}$)  & ($10^{35}$)  &  & & \\
  \noalign{\smallskip}
 \hline
 \noalign{\smallskip}
J08408
    & low            &  0.05          &$0.30_{-0.30}^{+0.16}$  &$0.44_{-0.09}^{+0.13}$ &                --                &                 --                  & 5.6  &0.078   & $1.7/57$ &  & \cr     
   & low            & 0.05           & $1.54_{-0.47}^{+0.50}$  &$1.02_{-0.25}^{+0.26}$ &$69_{-11}^{+14}$ & $(9.5_{-7.9}^{+56.0})\times10^{2}$ & 5.5   &0.076  & $1.18/55$&  $4.4\times10^{-5}$& \ref{sfxt10:fig:xrtspecfits1}	\cr   
    & very low\tablefootmark{d} & 0.009          & $0.30_{-0.30}^{+0.00}$ &$3.24_{-0.18}^{+0.18}$ &      --                &                 --                    & 0.09  & 0.001 & 1.47/194 &  	   &    \cr     
    & very low\tablefootmark{d} & 0.009          & $0.30_{-0.30}^{+0.11}$ &$1.98_{-0.32}^{+0.31}$ & $99_{-12}^{+13}$& $15 _{-5}^{+11}$  & 0.19   & 0.003  & 1.18/192 &  $6.9\times10^{-10}$ & \ref{sfxt10:fig:xrtspecfits1} \cr 
\noalign{\smallskip\hrule\smallskip}
 J16328  
    & low          &0.08      & $13.56_{-1.61}^{+1.82}$  & $1.35_{-0.26}^{+0.28}$&                --                &         --                                                     &16  &1.4  &$0.84/110$  &  & \ref{sfxt10:fig:xrtspecfits2} \cr 

     & very low\tablefootmark{e} &0.007    &$1.54_{-1.54}^{+0.99}$    &$0.30_{-0.22}^{+0.39}$  &                --                &         --                        &1.1    &0.06&1.66/242  &  & \cr 
    & very low          &0.007    &$4.37_{-1.50}^{+1.78}$     &$1.01_{-0.48}^{+0.52}$  & $46_{-13}^{+20}$ &  $(7.9_{-7.8}^{+30.0})\times10^5$ & 0.95 &0.06  &1.12/240 &  $3.1\times10^{-21}$ & \ref{sfxt10:fig:xrtspecfits2} \cr 
\noalign{\smallskip\hrule\smallskip}							      
 J16465
      & high        &$>$0.25        &$2.76_{-0.35}^{+0.39}$  &$1.05_{-0.15}^{+0.16}$&                --                &         --                                         &43 &9.7  &$1.02/108$ &  & \cr              
       & high        &$>$0.25        &$3.02_{-0.38}^{+0.43}$  &$1.13_{-0.16}^{+0.16}$ &$51_{-11}^{+32}$&$(2.3_{-2.3}^{+36.0})\times10^5$          &43 &9.8  &$0.93/106$ &  $7.5\times10^{-3}$ & \ref{sfxt10:fig:xrtspecfits3} \cr 
       & medium  &[0.15--0.25[  &$2.04_{-0.32}^{+0.37}$ &$0.90_{-0.15}^{+0.16}$ &                --                &         --                                       &23 &5.0  &$1.18/102 $ &  & \cr 
       & medium  &[0.15--0.25[  &$3.53_{-0.86}^{+0.98}$ &$1.28_{-0.24}^{+0.25}$ &$12_{-3}^{+3}$     &$(5.9_{-4.5}^{+19.0})\times10^{2}$     &22 &5.3  &$1.04/100$  & $1.8\times10^{-3}$ & \ref{sfxt10:fig:xrtspecfits3} \cr 
       & low         &$<$0.15         &$2.26_{-0.26}^{+0.28}$ &$1.42_{-0.14}^{+0.14}$ &                  --              &          --                                   &8.0& 1.8 &$1.32/120$ &  & \cr 
       & low          &$<$0.15        &$2.56_{-0.30}^{+0.32}$ &$1.52_{-0.15}^{+0.15}$ &$55_{-9}^{+11}$& $(3.4_{-2.8}^{+15.0})\times10^{4}$      &7.9 &1.8 &$1.17/118$ &  $8.1\times10^{-4}$ & \ref{sfxt10:fig:xrtspecfits3} \cr 
  \noalign{\smallskip}
 \hline
  \end{tabular}
  \end{center}
\tablefoot{
Uncertainties are given at 90\,\% confidence level for one interesting parameter. 
\tablefoottext{a}{Average observed 2--10\,keV fluxes in units of 10$^{-12}$~erg~cm$^{-2}$~s$^{-1}$.}
\tablefoottext{b}{Average 2--10\,keV  luminosities in units of 10$^{35}$~erg~s$^{-1}$ calculated by adopting the distances of Table~\ref{sfxt10:tab:dutycycle}.
}
\tablefoottext{c}{   F-test probability for the addition of the blackbody component (previous line).  }
\tablefoottext{d}{Fit performed with a column density constrained to be larger than the one derived from optical extinction towards the optical 
counterpart (see Sect~\ref{sfxt10:spec_out_of_outburst}). }
\tablefoottext{e}{Fit performed with column density constrained to be  larger than the Galactic value (see Sect~\ref{sfxt10:spec_out_of_outburst}). }
}
\end{table*} 

\begin{figure}
\begin{center}
\hspace{-0.5truecm}
\centerline{\includegraphics[width=6.cm,angle=270]{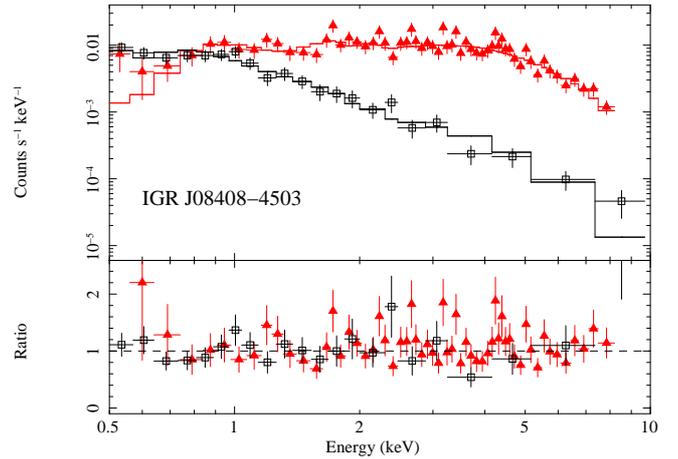}}
\caption{Spectroscopy of the 2011--2012 campaign on J08408.
            Red filled triangles and black empty squares mark low and very low states, respectively.  
             {\it Top panel}: XRT data fit with an absorbed power law and blackbody. 
             {\it Bottom panel}: the data/model ratio. 
 		\label{sfxt10:fig:xrtspecfits1} 
}
\end{center}
\end{figure}

       \subsection{Soft X--ray  inactivity duty cycle\label{sfxt10:idc}}

Our monitoring pace ensures a casual sampling of the X--ray light curve at a 
resolution of $\sim 3$--4\,d  over a $\sim 1$\,yr baseline, so we can follow the procedures
detailed in \citet{Romano2009:sfxts_paperV,Romano2011:sfxts_paperVI}, to
calculate the percentage of time each source spent in each flux state. 
\citet{Romano2009:sfxts_paperV,Romano2011:sfxts_paperVI} defined 
three states,
 {\it i)} BAT-detected flares, 
{\it ii)} intermediate state (all observations yielding a firm detection, outbursts excluded),  
{\it iii)} `non detections' (detections with a significance below 3$\sigma$) with exposure in excess 
of 900\,s (to account for  
non detections obtained during very short exposures due to our observations being 
interrupted by a higher figure-of-merit GRB event).  

The duty cycle of {\it inactivity} is defined \citep{Romano2009:sfxts_paperV}  as 
the time each source spends {\it undetected} down to a flux limit of 
1--3$\times10^{-12}$ erg cm$^{-2}$ s$^{-1}$,  
\begin{equation}
{\rm IDC}= \Delta T_{\Sigma} / [\Delta T_{\rm tot} \, (1-P_{\rm short}) ] \, ,  
\end{equation}
where  
$\Delta T_{\Sigma}$ is the sum of the exposures (each longer than 900\,s) accumulated in all observations 
where only a 3$\sigma$ upper limit was achieved 
(Table~\ref{sfxt10:tab:dutycycle},  Col.~5), 
$\Delta T_{\rm tot}$ is the total exposure accumulated (Table~\ref{sfxt10:tab:campaign}, Col.~6), 
and $P_{\rm short}$ is the fraction of time lost to short observations  
(exposure $<900$\,s, Table~\ref{sfxt10:tab:dutycycle}, Col.~6). 
The flux limits 1--3$\times10^{-12}$ erg cm$^{-2}$ s$^{-1}$ 
(Table~\ref{sfxt10:tab:dutycycle}, Col.~3) are obtained by converting 
the limiting count rates (derived from a measurement of the local background during the whole
campaigns) with a count rate to flux conversion derived from the best fit models of the 
`low' (J08408 and J16328) and `medium' (J16465) 
spectra in Table~\ref{sfxt10:tab:xrtspecfits} (Sect.~\ref{sfxt10:spec_out_of_outburst}). 
For the initial sample we recalculated the values in \citet[][]{Romano2011:sfxts_paperVI} 
based on the whole length of the campaigns. 
For the orbital monitoring sample we adopted the best fit to the first sequence  
described in \citet[][]{Romano2012:sfxts_16418} for J164182, 
the total spectrum in \citet[][]{Ducci2013:sfxts_17354}  for J17354, 
and the `medium' spectrum in \citet[][]{Romano2010:sfxts_18483} for 
J18483. 
Table~\ref{sfxt10:tab:dutycycle} also reports the limiting luminosities (Col.~4)
and the distances adopted (Col.~9). 

For the new sample we obtain that ${\rm IDC} = 67$, $61$, and $5$\,\%, 
for J08408, J16328, and  J16465, respectively
(Table~\ref{sfxt10:tab:dutycycle}, Col.~7).

\subsection{Out-of-outburst X--ray spectroscopy}  \label{sfxt10:spec_out_of_outburst}

Let us now consider the emission outside the bright outbursts. 
For J08408, J16328, and J16465 (totalling about $1800$, $3000$, and $7600$ counts, respectively)
we extracted the events in each observation when a detection was achieved 
(point {\it ii)} in Sect.~\ref{sfxt10:idc}), 
thus effectively selecting an intermediate, non quiescent state, and accumulated the mean spectrum.
For IGR~J08408$-$4503 and IGR~J16328$-$4726 we name this spectrum `low' 
(see Table~\ref{sfxt10:tab:xrtspecfits}).
For J16465, we split the events in the ranges $<0.15$ counts s$^{-1}$ (`low'), 
$0.15$--$0.25$  counts s$^{-1}$ (`medium'), and $>0.25$  counts s$^{-1}$ (`high'). 

Furthermore, 
we accumulated all data for which no detections were obtained as single exposures 
(point {\it iii)} in Sect.~\ref{sfxt10:idc},  whose combined exposure is $\Delta T_{\Sigma}$)
and extracted spectra (`very low' in Table~\ref{sfxt10:tab:xrtspecfits}, 
$\sim 300$--$500$ counts each), we binned them to 1 count bin$^{-1}$,   
and used Cash statistics\footnote{See, 
\href{http://www.swift.ac.uk/xrt\_spectra/docs.php}{http://www.swift.ac.uk/xrt\_spectra/docs.php} .}
for the fitting.
On these event lists, we performed a detection, and the resulting cumulative mean count 
rates  are reported in Table~\ref{sfxt10:tab:dutycycle} (Col.~8).

For all event lists exposure maps and ARFs were created as detailed in 
\citet{Romano2009:sfxts_paperV}. 
The spectra were rebinned with a minimum of 20 counts per energy bin, and 
fit in the 0.5--10\,keV (J08408 and J16328) and  0.3--10\,keV (J16465) energy ranges
with a simple absorbed power-law model, and an absorbed power-law model plus a 
blackbody ({\sc bbodyrad}) when the residuals indicated a soft X-ray excess. 
In that case, the F-test probability for the addition of such component is reported 
in Table~\ref{sfxt10:tab:xrtspecfits} (Col.~11) along with the fit parameters (Cols.~4--7)
and their 90\,\% confidence level (precision) errors for one interesting parameter. 
We note that, given the relatively poor statistics in the soft X--ray, this thermal component is
to be considered a convenient parameterization of the soft excess, 
rather than the modelling of a physical feature.

\begin{figure}
\begin{center}
\hspace{-0.5truecm}
\centerline{\includegraphics[width=6.cm,angle=270]{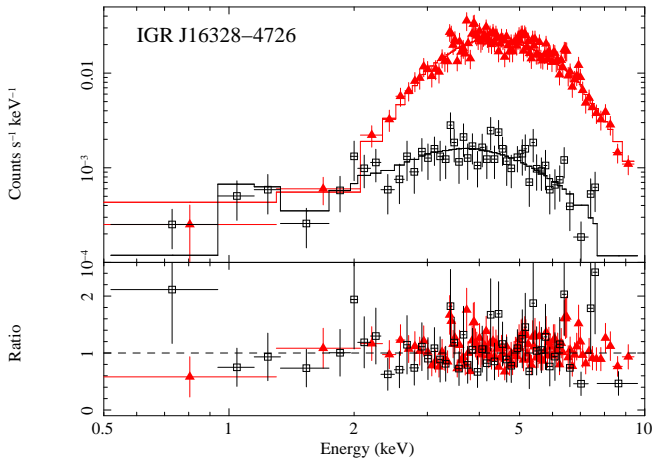}}
\caption{Spectroscopy of the 2011--2013 campaign on J16328.
            Red filled triangles and black empty squares mark low and very low states, respectively.  
            {\it Top panel}: XRT data fit with an absorbed power law (low) and absorbed power law plus blackbody (very low),
            respectively. 
            {\it Bottom panel}: the data/model ratio.
 		\label{sfxt10:fig:xrtspecfits2} 
}
\end{center}
\end{figure}

Figure~\ref{sfxt10:fig:xrtspecfits1} shows the best-fit spectra of  J08408. 
The  `low'  spectrum of J08408 is characterised by the presence of 
a soft ($<2$\,keV) component in excess of a simple absorbed power-law continuum, 
as demonstrated by the trend for the absorption column density to peg at an 
unphysical null value. 
Even when the absorbing column is constrained to be above the value 
derived from the optical extinction towards the optical counterpart
HD~74194, 0.3$\times10^{22}$\,cm$^{-2}$ \citep[][]{Leyder2007},  
in excess of the Galactic value of 0.932$\times10^{22}$\,cm$^{-2}$
\citep[][]{LABS}, our fit is 
formally unacceptable (`low' in Table~\ref{sfxt10:tab:xrtspecfits}). 
Although the addition of a thermal component improves the fit ($p=4.4\times10^{-5}$), 
due to the low statistics below $\sim2$\,keV, the black-body radius is not well constrained.  
Our results are consistent with those of 
\citet[][a 26\,ks {\it XMM-Newton} observation, col.~4 in table~3]{Bozzo2010:quiesc1739n08408}. 
For the `very low' spectrum we obtain spectral parameters consistent with those 
of the fainter spectrum in \citet[][col.~2 in table~3]{Bozzo2010:quiesc1739n08408}, 
although our $N_{\rm H}$ pegged to the Galactic value and 
our derived $R_{\rm BB}$ is significantly smaller.

Figure~\ref{sfxt10:fig:xrtspecfits2} shows the spectra of J16328. 
The `low' spectrum  is fit well by a simple absorbed power law, and 
is a factor of 3 fainter than the lowest state observed in the 2009 June 10 
outburst \citep[][]{Romano2013:Cospar12}, and shows consistent spectral parameters. 
The addition of a soft component is required only for the `very low' spectrum, 
but the blackbody radius is unconstrained.
The spectra outside of outburst are relatively softer than those observed during the
bright outbursts,  
as is generally observed for SFXTs when fitting the soft X-ray band data, 
only \citep{Romano2013:Cospar12}. 
Our `low' spectrum results are consistent with those of 
\citet[][a 22\,ks {\it XMM-Newton} observation]{Bozzo2012:HMXBs}.

\begin{figure}
\begin{center}
\hspace{-0.5truecm}
\centerline{\includegraphics[width=6cm,angle=270]{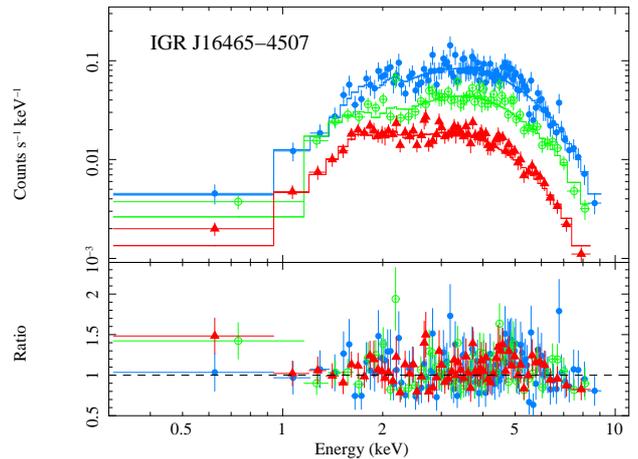}}
\caption{Spectroscopy of the 2013 campaign on  J16465.
            {\it Top panel}: XRT data fit with an absorbed power law plus blackbody.  
            {\it Bottom panel}: the data/model ratio.
            Filled blue circles, green empty circles, and red filled triangles
            mark high, medium, and low states, respectively. 
 		\label{sfxt10:fig:xrtspecfits3} 
}
\end{center}
\end{figure}

Figure~\ref{sfxt10:fig:xrtspecfits3} shows the spectra of  J16465. 
While the `high' spectrum is fit adequately by a simple absorbed power law, 
the residuals still show a trend for an extra soft component, so further fits were performed 
with the addition of blackbody component. 
The `low' spectrum has flux comparable to that of the spectrum observed by 
\citet[][{\it Suzaku} observation]{Morris2009:suzaku} for which consistent values of 
absorbing column and photon index were found.

\begin{figure*}
\begin{center}
\vspace{-4truecm}
\centerline{\includegraphics*[angle=0,width=19cm]{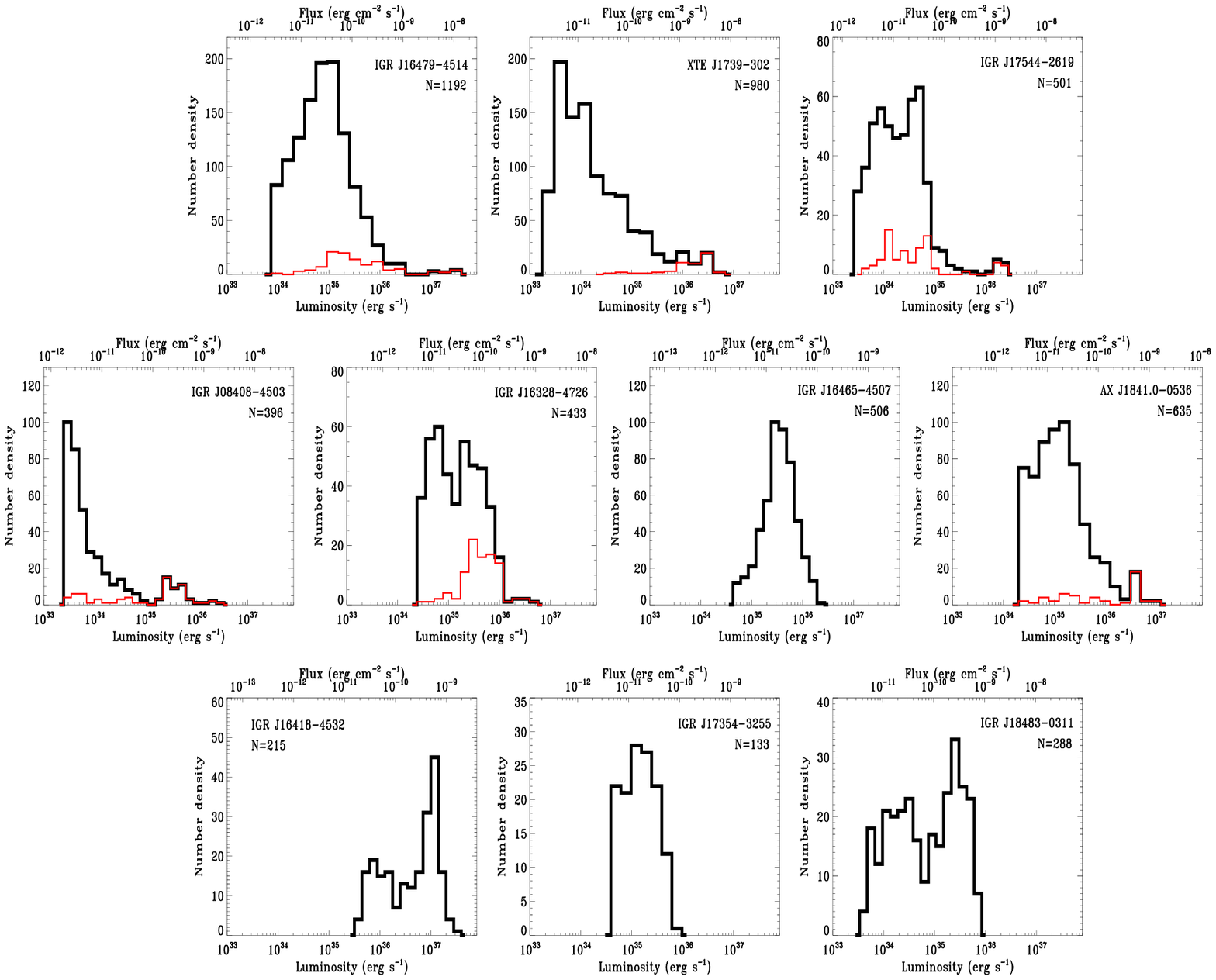}}

\vspace{-7truecm}
\end{center}
\vspace{-1.5truecm}
\caption{
Distributions of the XRT 2--10\,keV  luminosity (lower axis) and flux (unabsorbed, upper axis)  
drawn from the light curves binned at 100\,s.
{\it Top}: SFXTs monitored for two years \citep[][]{Romano2011:sfxts_paperVI}; 
the thin red histograms 
show the part of the data collected as outburst observations
thus including both the initial bright flare and the follow-up observations. 
{\it Middle}: SFXTs monitored for one year \citep[][]{Romano2009:sfxts_paperV}; 
the thin red histograms  show outburst observations collected outside 
of the monitoring campaign (one outburst per source). 
{\it Bottom}: SFXTs monitored for one orbital period 
\citep[][]{Romano2010:sfxts_18483,Ducci2013:sfxts_17354,Romano2012:sfxts_16418}.
The sample size, N, is reported in each panel. 
}
\label{sfxt10:fig:nhistos3}
\end{figure*}

\begin{figure*}
\begin{center}
\vspace{-7truecm}
\centerline{\includegraphics*[angle=0,width=19cm]{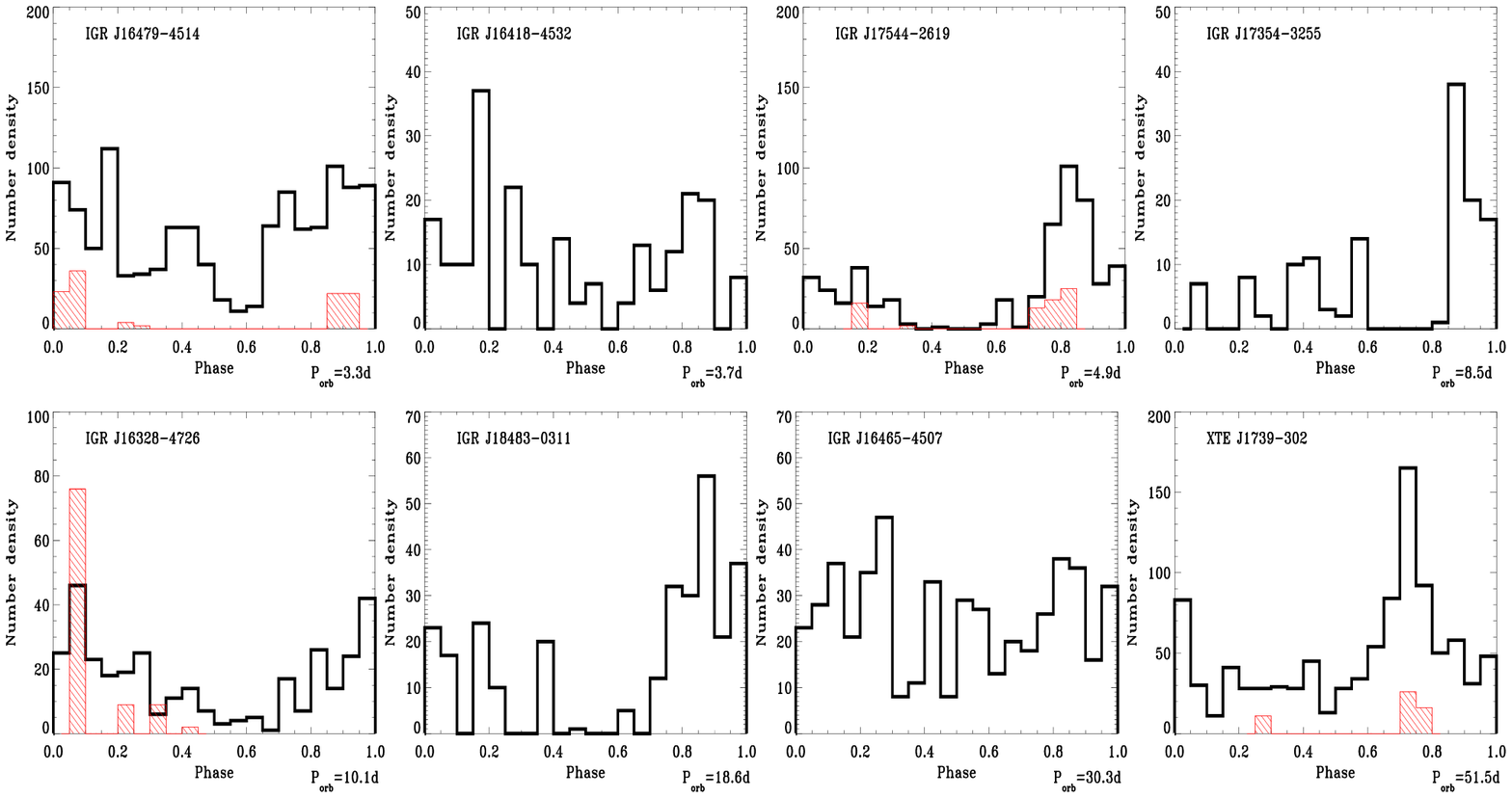}}
\end{center}
\vspace{-10.5truecm}
\caption{Distribution of the XRT count rates  (0.3--10\,keV) folded at the orbital periods, 
with the sources ordered by orbital period. Color coding is the same as in Fig.~\ref{sfxt10:fig:nhistos3}.  
}
\label{sfxt10:fig:histos2}
\end{figure*}

\subsection{Count rate,  flux, and luminosity distributions}  \label{sfxt10:cr_distros}
Following the procedures detailed in \citet[][]{Romano2011:sfxts_paperVI}, 
we calculate the percentage of time 
J08408, J16328, and  J16465 spend at a given flux state.  
To also place their behaviour in a broader context, 
we applied the same procedure for both the newly processed data on the initial  
and for the  orbital monitoring samples. 

Figure~\ref{sfxt10:fig:nhistos3} (black solid lines) shows the differential 
distributions of the 2--10\,keV luminosities, drawn from the XRT light curves binned at 100\,s, 
after removal of the observations where a detection was not achieved, 
ordered by length of monitoring campaign. 
Since the uncertainty in this conversion is dominated by those on the distance 
determinations, Fig.~\ref{sfxt10:fig:nhistos3} also reports the 
flux scale (2--10\,keV, unabsorbed), as the top x-axis. 
The conversion factors for the initial monitoring sample were those 
calculated in \citet[][]{Romano2011:sfxts_paperVI}; for the orbital monitoring sample (see Sect.~\ref{sfxt10:olddata}) 
those calculated in \citet[][]{Romano2010:sfxts_18483,Romano2012:sfxts_16418} and \citet[][]{Ducci2013:sfxts_17354}. 
For the new monitoring sample, the conversion factors were derived from the spectroscopy 
in Table~\ref{sfxt10:tab:xrtspecfits}.
The first row shows the reanalyzed data on the 
three SFXTs monitored for two years \citep[][]{Romano2011:sfxts_paperVI}.  
We distinguish, among the data, those that were taken during an outburst 
(2 for J16479, and 3 for J1739 and  J17544) 
as a thin red 
histogram. 
We note how the outburst data have one bright peak in the range 
$\sim 10$--$70$\,counts\,s$^{-1}$ 
(corresponding to a few $10^{36}$--$10^{37}$\,erg\,s$^{-1}$), 
while the remainder spreads beneath the main peak of the luminosity distribution. 
This is due to the way the data were collected, as a BAT trigger follow-up, 
hence with a statistically very rich first orbit of data 
sometimes followed by  an intense monitoring up to $\sim10$\,ks per day  
until the source went back to the pre-outburst levels. 
Figure~\ref{sfxt10:fig:histos2}, which shows the count rate (CR) distributions in phase\footnote{The 
XRT light curves were first phased at the known periods, then histogrammes  
were created.}  (periods in Table~\ref{sfxt10:tab:dc13}, Col.~2), 
further illustrates this in the panel on J16328: 
the hashed  histogram data were taken consecutively after a bright 
outburst that was followed intensively. 

The second row of Fig.~\ref{sfxt10:fig:nhistos3} shows the 4 sources that
were monitored for one year and never went into outburst while being monitored. 
To asses the overall distributions, 
we therefore selected one outburst 
and added the data as a thin red histogram.
J16465 never triggered the BAT, so no data were added. 
For J08408 only the tail of a distribution probably peaking well below the
XRT sensitivity at this binning is seen, but the outburst data clearly map another distribution, 
with a peak at a few counts\,s$^{-1}$ and extending up to about $\sim50$\,counts\,s$^{-1}$,
corresponding to a few $10^{35}$--$10^{36}$\,erg\,s$^{-1}$. 
Similarly, J1841 shows a non-outburst  distribution peaking at 
about $\sim 0.1$\,counts\,s$^{-1}$, while the outburst data peak at a few counts\,s$^{-1}$.
J16328 shows a non-outburst distribution qualitatively similar to the one observed in
J1739 and J1841, but the statistics do not allow us to determine whether
the outburst data fall on the tail of the main distribution or if they can be distinguished from it. 
Based on these findings, 
the most probable X-ray flux for J08408 is $\la 2\times10^{-12}$ erg cm$^{-2}$ s$^{-1}$ (2--10\,keV, unabsorbed),  
for J16328 is $\sim 10^{-11}$ erg cm$^{-2}$ s$^{-1}$. 
These are about two orders of magnitude lower than the bright outbursts for these two sources. 
J16465 shows a well defined distribution peaking at $\sim 0.1$\,counts\,s$^{-1}$,
corresponding to $\sim 2\times10^{-11}$ erg cm$^{-2}$ s$^{-1}$.

The third row of Fig.~\ref{sfxt10:fig:nhistos3} shows the  distributions for the 
three SFXTs monitored for one orbital period, none of which has XRT outburst data available. 
We note that J18483 triggered the BAT once on 2008 August 4, but no XRT data are available.
Furthermore, while J16418 triggered the BAT four times since \sw's launch, one trigger did not have 
XRT follow-up, two were sub-threshold (and showed a light curve peaking at $\sim 5$ counts s$^{-1}$),
and the last had a very late follow-up, so no data with matching quality to those of the remainder of 
the sample are available. 
These data need to be taken with caution, as they were collected with an entirely different observing 
strategy. Indeed, while the yearly monitoring is a casual sampling of the light curves with few points 
per period, these observations were collected with an intensive campaign during one or few orbital periods. 
Therefore the effects of short timescale variability 
\citep[variations of one order of magnitude are quite 
common, see][]{Romano2010:sfxts_18483,Romano2012:sfxts_16418,Ducci2013:sfxts_17354} 
may play the dominant role in this case.

The CR distributions in phase (Fig.~\ref{sfxt10:fig:histos2}) 
match reasonably well (considering the lower S/N in the XRT data) 
the BAT light curves folded at the orbital periods, as shown in 
\citet[][]{Romano2014:sfxts_catI}, 
although the eclipse throughs expected in J16479 and J16418 are less deep. 
This is a common occurrence in eclipsing HMXBs, and 
is discussed in terms of dust scattering by \citet[][]{Bozzo2008:eclipse16479} 
and by \citet[][]{Drave2013:16418} for J16418 as reprocessing 
of the intrinsic neutron star emission by the supergiant  dense wind.

\section{Discussion\label{sfxt10:discussion}}

\subsection{Soft X--ray long term properties: J16465 is not an SFXT \label{sfxt10:disc_monitoring}}
In this paper we report the results of a \sw/XRT monitoring of 
J08408, J16328, and J16465 
along a baseline of over two years and place them in the broader context of the 
SFXT sample.

During the campaigns only J16328 triggered the BAT 
and the properties of this bright flare, $\Gamma_{\rm 2011}=3\pm1$, 
$F_{\rm 20-50\,keV} = 2.8\times10^{-10}$ erg cm$^{-2}$ s$^{-1}$, 
are consistent with those observed during the only other outburst recorded by 
\sw\ on this source ($\Gamma_{\rm 2009}=2.6\pm0.4$, 
$F_{\rm 20-50\,keV} = 7.1\times10^{-10}$ erg cm$^{-2}$ s$^{-1}$). 
Given the lack of observed outbursts during our monitoring, 
and considering the outburst history of the three sources, 
we estimate that they spend less that 1\,\% of their time in bright outbursts.

The main purpose of our monitoring is to exploit the unique flexibility 
of \sw\ to continue our characterisation of the long-term behaviour and 
emission outside the bright outbursts in SFXTs. 
J08408 and J16328 show activity
at a level of 1--2 orders of magnitude lower than the bright outbursts,
as previously observed for the initial monitoring sample 
\citep[][and references therein]{Romano2011:sfxts_paperVI}. 
Figure~\ref{sfxt10:fig:xrtlcvs} shows that, when the data are binned to a daily timescale, 
the dynamical range  (25--50) of these two SFXTs is somewhat smaller than that of 
the initial sample that instead showed variations spanning more than two-orders of magnitude. 
Nevertheless, when we take into account the bright outbursts of J08408 
and J16328 and the deep 3$\sigma$ upper limits obtained combining all 
non-detections, their DR increases to 7400 and 750, respectively, 
typical of the SFXT population. In either case, however, they 
do not reach the four orders of magnitude observed in the initial sample
\citep[fig.~1 in ][]{Romano2011:sfxts_paperVI}. 

The intermediate state of emission we observed from these sources during our monitoring 
is characterised by non-thermal emission (hence accretion onto the compact object) 
following the previously observed harder-when-brighter trend \citep[e.g.][]{Romano2011:sfxts_paperVI},  
as well as by a soft excess whose strength becomes dominant in 
the`very low' spectra. 
We note that the addition of thermal components similar to the ones observed in 
other HMXBs \citep[e.g.][]{Hickox2004,vanderMeer2005}, 
and in particular in J08408 \citep{Bozzo2010:quiesc1739n08408}, 
improves the fit but, due to the low statistics below $\sim 2$\,keV, 
the parameters are often poorly constrained or unconstrained. 

Our observations of J08408 and J16328 show that 
this intermediate state is characterised by 
soft X--ray flux variability observed on timescales of a few hundred seconds, 
as  also observed in the initial and orbital 
monitoring samples, which is generally explained in terms of the clumpiness of the 
wind of the supergiant companion \citep[e.g.][]{Walter2007}. 

\setcounter{table}{7}  
 \begin{table}  
 \tabcolsep 2pt         
 \begin{center}         
 \caption{Duty cycles as a function of orbital periods. }
 \label{sfxt10:tab:dc13}       
 \begin{tabular}{lrrrrrr} 
 \hline 
 \hline 
\noalign{\smallskip}
  Name                        & Orbital &XRT   &XRT      & IGR   & IGR    & Ref. \\
                                   & Period&IDC\tablefootmark{a}     & DC\tablefootmark{b}  & DC\tablefootmark{c}    &DC\tablefootmark{d}  & $P_{\rm orb}$\\
                                   & (d)& (\%)  & (\%)  &(\%)    &(\%)     & \\  
 \noalign{\smallskip}
 \hline
 \noalign{\smallskip} 
IGR~J16479$-$4514     &$3.3193$   &  19   &   1.8 &  2.8      &2.39  &   1 \\ 
IGR~J16418$-$4532    &$3.73886$ & 11    &  26 &  1.3      &0.90 &   2 \\ 
IGR~J17544$-$2619     &$4.926$     & 55    &  0.7 &  0.51    &0.72   &   3 \\ 
IGR~J17354$-$3255    &$8.448$     & 33    &  0.0 &  --       &--    &   4 \\   
IGR~J16328$-$4726    &$10.076$   & 61   &  6.3 &  --       &--    &   5 \\  
IGR~J18483$-$0311    &$18.545$   & 27   &  12 &  3.2      &3.31  &   6 \\ 
IGR~J16465$-$4507    &$30.243$   & 5     &  0.0 &  0.24    &0.13  &   7 \\  
XTE~J1739$-$302       &$51.47$    & 39   &  4.1 &  0.83   &0.89  &  8 \\  
AX~J1841.0$-$0536      & --            & 28   &  8.7 &  0.49    &0.44  &  -- \\  
IGR~J08408$-$4503    & --            & 67   &  3.8 &  --       &0.16 &  -- \\   
  \noalign{\smallskip}
  \hline
  \end{tabular}
  \end{center}
\tablefoot{
\tablefoottext{a}{Inactivity duty cycle in the 0.3--10\,keV band.}
\tablefoottext{b}{XRT  duty cycle at the \inte{} sensitivity for each object (Sect.~\ref{sfxt10:disc_dc13}).}
\tablefoottext{c}{Derived from \citet[][]{Ducci2010}, in the 20--40\,keV band.}
\tablefoottext{d}{From \citet[][]{Paizis2014}, in the 17--30\,keV band.}
}
\tablebib{
(1) \citet{Romano2009:sfxts_paperV}; 
(2) \citet{Drave2013:16418};  
(3) \citet{Clark2009:17544-2619period}; 
(4) \citet{Dai2011:period_17354}; 
(5) \citet{Corbet2010:16328-4726}; 
(6) \citet{Levine2006:igr18483}; 
(7) \citet{LaParola2010:16465-4507_period}; 
(8)  \citet[][]{Drave2010:17391_3021_period}. 
}
  \end{table}

For J16465, \citet[][]{LaParola2010:16465-4507_period} 
reported that the BAT data show a narrow DR 
($<10$ in the 15--50\,keV band) and no flaring activity, 
and suggested this source is a faint supergiant HMXB, 
probably fed by a rather homogeneous wind, 
as opposed to a SFXT. 
\citet[][]{Romano2014:sfxts_catI} also report a very scarce activity (detections
in excess of 5$\sigma$) in the first 100 months of BAT data,  
and no outbursts in 9 years of \sw\ operations. 
However, more outbursts were reported by 
\citet[][]{Clark2010:16465-4507_period2},
who instead classify this source as intermediate, and by \citet[][]{Ducci2010}. 
Bearing this in mind, and in light of our soft X--ray 
findings, we discuss J16465 separately from the other two sources in the 
new monitoring sample.  
The results of the intensity-selected soft X--ray spectroscopy of this source 
(Table~\ref{sfxt10:tab:xrtspecfits}) can be directly compared with those on 
J16479 \citep[table~8 of][]{Romano2011:sfxts_paperVI}
that shows comparable luminosity levels in the high, medium and low spectra. 
In particular, we note both the consistency of the photon indices, when a simple absorbed 
power-law model is adopted, and the general trend for harder-when-brighter emission,
as commonly observed in SFXTs. 
From this point of view, then, the spectral behaviour of J16465 is 
consistent with what we expect from the SFXT (as well as, of course, from 
the general sgHMXB) population.

However, 
the XRT overall DR is below 40, as typical of the general 
HMXB population, rather than of SFXTs, and very little variability is observed on
data binned at timescales of 100\,s, for which a  DR$\la  5$
is observed within one orbit. 
Furthermore, the histogram of the observed CR is single-peaked that, 
differently from the other SFXTs, does not show a secondary peak corresponding  to 
the outburst data.  The steepness of the wings of the distribution indicates that no emission 
is observed in excess of $\sim 1$\,counts\,s$^{-1}$. The full-width at zero intensity of the distribution is 
considerably less than 2 decades, while the other SFXTs exceed three orders of magnitude. 
%
Finally, the measured IDC (5\,\%) is at the very lower end  
of the observed distribution in SFXTs,
since the lowest value is that of J16418.  
To all intents and purposes, especially in consideration of the fact that 
this source is not particularly absorbed and its distance is at the high end of the SFXT distribution, 
J16465 is a persistent source in the XRT. 
The current soft X--ray data seem to point toward a Vela X-1-like source 
\citep[][]{Kreykenbohm2008}, as opposed  to an SFXT. 

We note that the discordant behaviour of J16465 with respect to 
that of the remainder of the \inte\ SFXT sample 
was also reported by \citet[][]{Lutovinov2013:HMXBpop}. 
Within their proposed model for wind-fed HMXBs hosting a neutron star, 
they produced a theoretical hard X--ray luminosity vs. orbital period 
($P_{\rm orb}$--$L_{\rm X}$) diagram to which both `normal' wind-fed HMXBs 
and SFXTs are compared. 
While  normal wind-fed HMXBs are observed to lie above the lower limit of luminosity 
allowed  for a given period ($L_{\rm X} \sim P_{\rm orb}^{-4/3}$),  
the SFXT population shows median luminosity beneath this curve.  
Therefore, the flaring observed in SFXTs can be explained within this context, provided 
that some mechanism,  such as magnetic arrest, inhibits accretion.  
In the $P_{\rm orb}$--$L_{\rm X}$ diagram, however, IGR J16465$-$4507 lies in the 
same allowed area as normal wind-fed HMXBs. 
We note that our luminosity distribution (Fig.~\ref{sfxt10:fig:nhistos3}) for this source reaches down to
$\ga 4\times10^{34}$ erg s$^{-1}$, 
and the XRT lowest detection (obtained by summing up all 3-$\sigma$ upper limits, 
see Sect~\ref{sfxt10:xrtlcvs}) corresponds to $\sim 3.8\times10^{34}$ erg s$^{-1}$.
These values are much closer to the $L_{\rm X} \sim P_{\rm orb}^{-4/3}$ limit than the 
\inte\ data, so it is possible that deeper observations of this source 
and a better determination of the distance\footnote{Note that the luminosity values 
become $\sim 2\times10^{34}$ erg s$^{-1}$ at the  distance of 9.4\,kpc 
adopted by \citet[][]{Lutovinov2013:HMXBpop}.} 
might just make it cross out of the allowed ranges in the $P_{\rm orb}$--$L_{\rm X}$ diagram,  
which occur at $\sim 10^{34}$ erg s$^{-1}$ for a period of about 30\,d 
(see fig.~10 of \citealt[][]{Lutovinov2013:HMXBpop}), like the remainder of the SFXT sample. 
If that were not the case, however, then this would add to the evidence that 
IGR J16465$-$4507 is indeed a normal wind-fed HMXB, as opposed to a SFXT.

\subsection{Duty cycles and orbital geometry\label{sfxt10:disc_dc13}}

A long-standing question in the SFXT field is whether the duty cycle is related to the orbital 
parameters, the period {\it in primis}. 
If the dominant source of X--ray variability in SFXTs were 
the properties of the binary geometry and inhomogeneity
of the stellar wind from the donor star, as proposed in the clumpy wind models
\citep[e.g.][]{zand2005,Negueruela2008,Walter2007}, then 
we could expect generally larger IDCs for larger orbital periods.  
We can now address this question with high-sensitivity data.  

We defined the {\it inactivity} duty cycle \citep[][]{Romano2009:sfxts_paperV} 
as the time each source spends undetected down to a flux limit of 
1--3$\times10^{-12}$ erg cm$^{-2}$ s$^{-1}$ (see Table~\ref{sfxt10:tab:dutycycle}),
thus exploiting the higher XRT sensitivity when compared with hard X--ray detectors 
(\inte{} IBIS/ISGRI\footnote{ISGRI reaches a sensitivity \citep{Paizis2013:GOLIA} of 20\,mCrab 
in the 17--60\,keV band, at the 5$\sigma$ level for 1 pointing ($\sim 2$\,ks).}
or \sw/BAT\footnote{BAT reaches a sensitivity \citep[][]{Krimm2013:BATTM}
of 12.9\,mCrab in the 15--50\,keV band, at 1$\sigma$ level for 1 orbit ($\sim 1$\,ks).}) 
and the regular sampling of our monitoring campaigns. 
The initial sample showed that these sources were actually active for the great 
majority of time when inspected at fluxes as low as those achievable with the high 
sensitivity of XRT.  Similarly, our reanalysis of the data on the orbital monitoring 
sample shows very low IDCs (11 to 33\,\%). 
The IDCs of  J08408 and J16328 (67 and 61\,\%, respectively) are by far the 
highest of the SFXT sample, as these sources are not detected for the majority of time. 
On the contrary, J16465 has an IDC of 5\,\%, 
which is consistent with the source being persistent.  

\begin{figure}
\begin{center}
\vspace{-0.5truecm}
\centerline{\includegraphics*[angle=0,width=9.3cm]{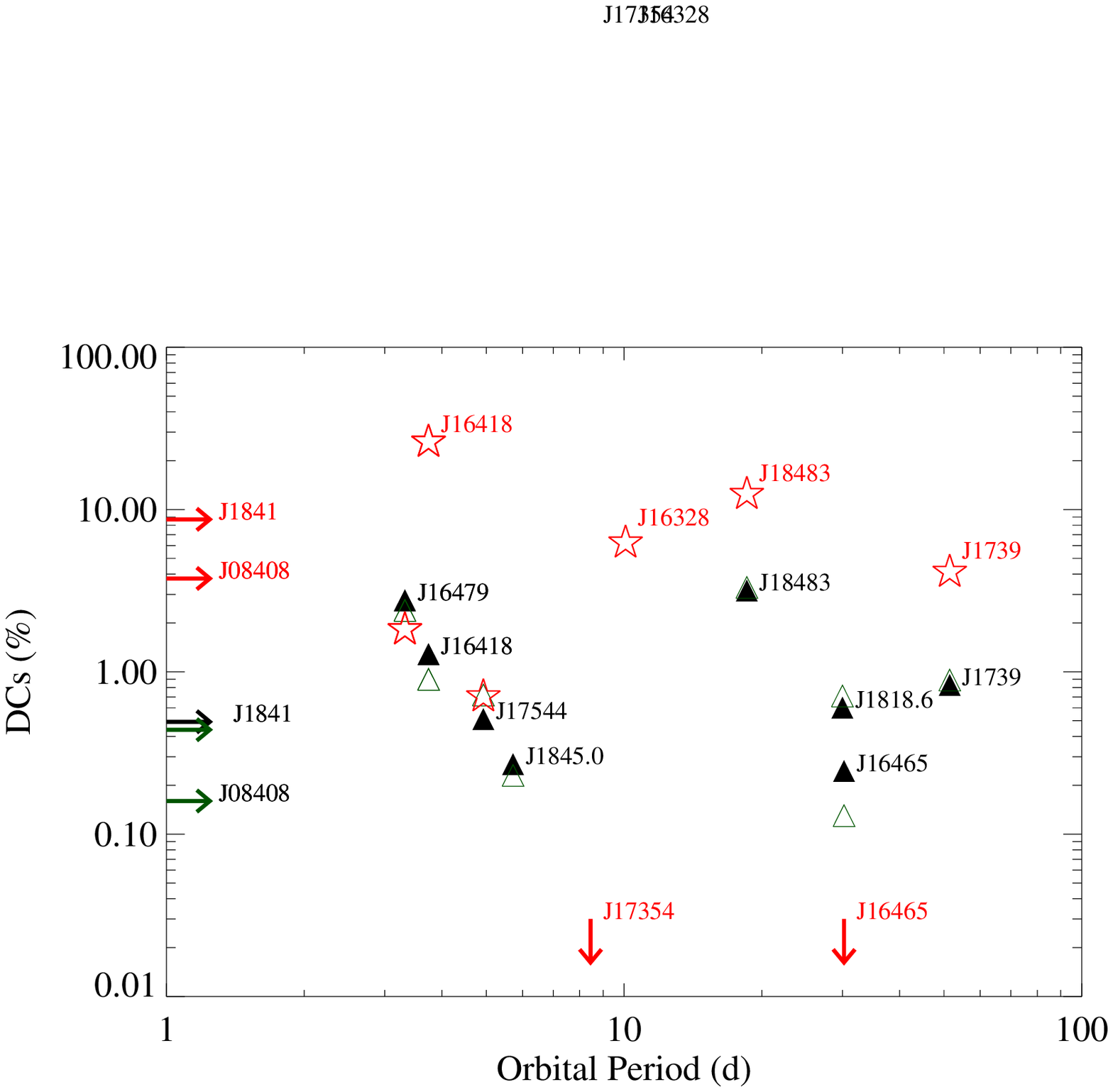}}
\end{center}
\vspace{-0.5truecm}
\caption{\inte-based duty cycles from \citet[][black filled triangles]{Ducci2010} and \citet[][green empty triangles]{Paizis2014}, 
and XRT duty cycle at the \inte{} sensitivity (red empty stars, Sect.~\ref{sfxt10:disc_dc13}). 
The  downward pointing arrows are consistent with 0. 
The right pointing arrows at $P_{\rm orb}=1$ are for sources lacking orbital period. }
\label{sfxt10:fig:porb_igrdcs}
\end{figure}

It is interesting to compare our IDC with the DC estimated from \inte, 
whose instruments have a lower sensitivity for fainter luminosity states of 
the SFXTs but which can provide longer-term observations.
The \inte{} IBIS/ISGRI data are presented in \citet[][7 objects in common with our sample; 
Table~\ref{sfxt10:tab:dc13}, Col.~5]{Ducci2010}, 
for which the most active sources are J18483 and J16479, and the least active is J16465. 
In \citet[][]{Ducci2010} the duty cycle (\inte{} DC) is defined as the ratio of the time 
the sources are detected in excess of 5$\sigma$ and the total exposure time
in the 20--40\,keV band. 
Similar results are found when the \inte{} DCs are drawn from the recent work of 
\citet[][8 objects in common; Table~\ref{sfxt10:tab:dc13}, Col.~6]{Paizis2014}. 
Figure~\ref{sfxt10:fig:porb_igrdcs} shows them as as function of the orbital period. 
We note that our IDC is generally anti-correlated with the \inte{} DCs,
with the notable exception of J16465.

%

\begin{figure*}
\begin{center}
\vspace{-0.5truecm}
\centerline{\includegraphics*[angle=0,width=19.8cm]{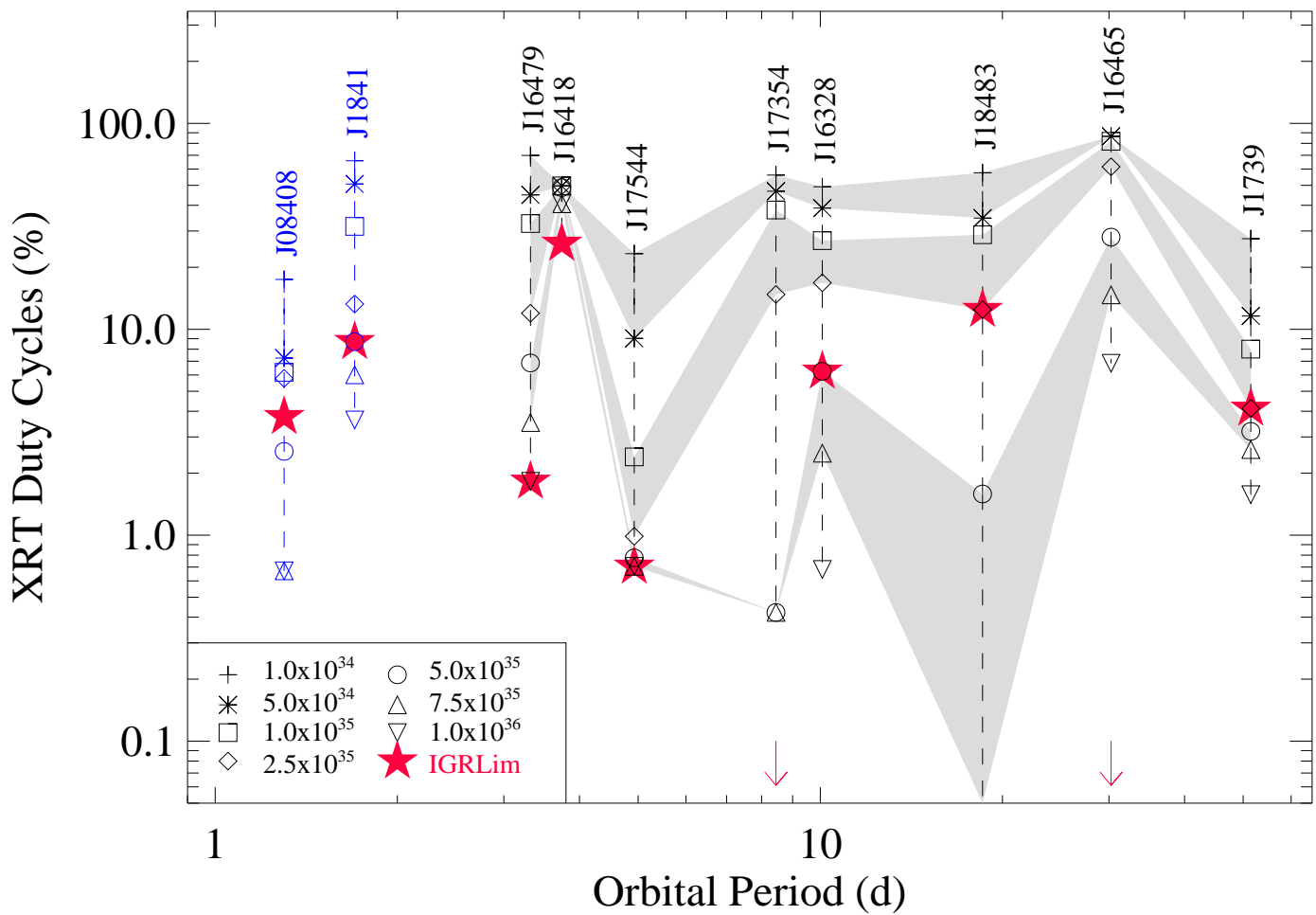}}
\end{center}
\vspace{-0.5truecm}
\caption{XRTDC (2--10\,kev) as a function of orbital period and for a range of 
2--10\,keV luminosities (see legend, in units of erg s$^{-1}$) in black.   
All values are also reported for J08408 and J1841, 
lacking an orbital period measurement, as blue data points at arbitrary orbital period below 2 days). 
Only points above 0.1\,\% are shown. 
The shaded areas mark the loci of XRTDC defined with contiguous 
luminosities.  
The red filled stars represent the XRT DC at the \inte{} sensitivity for each object
(the downward pointing arrows are consistent with 0).} 
\label{sfxt10:fig:porb_dc}
\end{figure*}

Both the XRT IDC and the \inte{} DC are based on the instrumental sensitivity
in the detector band. What follows is an attempt to overcome these biases. 
We define an {\it XRT luminosity-based duty cycle} (XRTDC) as the percentage of time
the source spends above a given luminosity,
 and we considered several luminosities in the range $L_{\rm 2-10\,keV}=10^{34}$--$10^{36}$ erg s$^{-1}$. 
Figure~\ref{sfxt10:fig:porb_dc} shows the XRTDC  as a function of the orbital period. 
We find that, clearly, the definition of duty cycle is strongly dependent on the luminosity 
assumed as lower limit for the calculation. 

In particular, we can also consider the XRTDC calculated for
the  luminosity corresponding to the \inte{} sensitivity for each object. 
We considered that IBIS reaches \citep{Paizis2013:GOLIA} 20\,mCrab (17--60\,keV) at the 
5$\sigma$ level for 1 pointing ($\sim 2$\,ks), 
and adopted the best \sw\ broad-band spectra obtained during outburst 
for each object, to convert from the IBIS band and the 2--10\,keV one.  
These points are also plotted in Fig.~\ref{sfxt10:fig:porb_dc} (red stars). 
The XRT DCs at the \inte{} sensitivity are reported in Table~\ref{sfxt10:tab:dc13}, Col.~4. 
They range from $\sim0.7$\,\% for J17544 to 26\,\%  for J16418 and 
there is a good match with the corresponding \inte{} values for J17544 and J16479.
At the \inte{} sensitivity J16465 and J17354 have a null DC, so all emission for these sources 
in below this threshold.

Once the different systematics coming into play in the different 
definitions of duty cycles are understood, we can consider once again the 
relationship between the duty cycle and the binary orbital period. 
We find that the SFXT duty cycles are not clearly correlated with the orbital period.
Therefore, wide orbits are not necessarily characterised by low duty 
cycles, as the clumpy wind models would predict. 
Instead, an intrinsic mechanism seems to be more likely responsible 
for the observed variability in SFXTs, i.e., either the 
wind properties or the compact object properties.

Finding it hard to justify radically different wind properties in SFXTs 
from those in `normal' HMXBs with the same companion spectral type, 
accretion inhibition mechanisms 
seem more plausible, especially in light of 
the very low DC  for  J17544 (as well as the other 
SFXT prototype J1739), for which \citet[][]{Bozzo2008} 
interpret the very large luminosity ranges observed on timescales as 
short as hours as transitions across the magnetic and/or centrifugal barriers.  
This is consistent with the conclusions of \citet[][]{Lutovinov2013:HMXBpop} 
that the flaring behaviour of SFXTs is likely related to the magnetic 
arrest of their accretion. 
Alternative mechanisms to partially inhibit accretion in HMXBs have 
been suggested by \citet[][]{Shakura2012:quasi_spherical} 
and applied so far to interpret the low luminosity regimes of a number of
classical supergiant X-ray binaries. 
The discussion of the applicability of their model to the SFXTs is beyond 
the scope of the present paper.

%

\subsection{Differential luminosity distributions\label{sfxt10:disc_distros}}

Other authors \citep[][]{Smith2012:FXRT_RXTE,Paizis2014} have 
used the longer baseline of relatively less sensitive \rxte\ and \inte\ 
data available--hence geared to best detect the bright flares--to construct 
cumulative luminosity distributions. 
In this paper we exploit the higher sensitivity XRT data to construct 
differential count rate (flux and luminosity) distributions, instead,
searching for faint features originating in different populations of flares
in the soft X--ray emission. 
We have indeed discovered that the SFXT prototypes, J1739  
and J17544, as well as J16479 and J08408, 
show two distinct populations of flares. 
The first one is due to the outburst emission and peaks (or reaches, as in the case of 
J08408) a few $10^{-9}$ erg cm$^{-2}$ s$^{-1}$. 
The second population is due to the out-of-outburst emission, which is characterised by 
emission spanning up to 4 orders of magnitude in DR (at 100\,s binning). 
While it is not possible to exclude that particular distributions of the clump and wind 
parameters may produce a double-peaked differential distribution,  
this behaviour is more 
easily explained in terms of different accretion regimes as predicted by the 
magnetic/centrifugal gating model or the quasi-spherical settling accretion model 
\citep[][]{Grebenev2007,Bozzo2008,Shakura2012:quasi_spherical,Shakura2013:off_states}.

\section{Summary and conclusions \label{sfxt10:conclusions}}

In this paper we have presented the first 
high-sensitivity (a few $10^{-12}$ erg cm$^{-2}$ s$^{-1}$) 
soft X--ray (0.3--10\,keV) long-term ($\ga 1$\,yr) 
monitoring with \sw/XRT of three relatively unexplored SFXTs, 
J08408, J16328, and J16465, which were chosen as those, 
among the SFXT sample, whose hard X--ray duty cycles are 
the lowest measured. We stress that our monitoring campaigns 
could only be performed thanks to the extraordinary flexibility 
in scheduling of \sw\ that makes such a monitoring effort cost-effective. 
Even though the single 1\,ks snapshots are shallow compared 
to the deep observations of the other pointed observations by \xmm\ or \suzaku, 
the advantages are many. 

First, thanks to the regular pacing, our data provide 
a casual sampling of the X--ray light curves at a 
resolution of $\sim 3$--4\,d  over a $\sim 1$\,yr baseline. 
They are therefore statistically representative of the long term properties of 
these sources that the long looks from other pointed telescopes can only 
sample, albeit more deeply, only rarely. 

Second, these data can be used to measure two defining quantities: 
{\it i)} the dynamical range, 
fundamental in discriminating between outbursts of classical supergiant 
HMXBs ($\lesssim 50$) and SFXTs ($\gtrsim100$), as described in 
\citet[e.g.][]{Negueruela2006:ESASP604,Walter2006};  and  
{\it ii)} the duty cycle as a function of the luminosity across the wide dynamical 
range spanned by the SFXTs, which is a measure of the activity of each source,  and 
for which different models for emission in SFXTs have contrasting predictions. 
We can also use the data to perform intensity-selected spectroscopy by combining
all short exposures, thus reaching the same intrinsic luminosities as those reached by 
the long looks, thus confirming their results. 

In this work, we not only created long term light curves, calculated dynamical ranges
and duty cycles,  spectroscopically studied the out-of-outburst emission, 
and created differential luminosity distributions for 3 new sources, 
but we also compared these properties with those of  the remainder of the SFXT sample. 
Our findings can be summarized as follows. 

\begin{itemize}
\item All SFXTs share out-of-outburst spectroscopic properties of non-thermal emission plus a soft excess
           (becoming increasingly more dominant as the source reaches the lowest emission states) 
           with the general population of supergiant HMXBs.  
           The spectroscopic investigation, therefore, is not  an efficient method of distinguishing SFXTs 
           within the HMXB sample. 
\item The behaviour of J08408 and J16328 resembles those of the SFXT prototypes: 
           the probable X-ray flux is about two orders of magnitude lower than their bright outbursts, 
           accounting for less than 1\,\% of the total time;   
           the overall dynamical range is DR$\sim 7400$ and $\sim 750$, respectively;   
           the IDC is $\sim 67$  and $61$\,\%, respectively, the highest in the SFXTs observed by XRT, 
           consistently with the hard X--ray observations.  
\item J16465 is to all intents and purposes a persistent source in the XRT, as opposed to an SFXT, 
          with its overall DR$\la 40$ and a duty-cycle of inactivity of  $5$\,\%.  
\item By examining the differential luminosity distributions of the SFXT sample, 
          we find that J17544,  J1739, J16479, and J08408, 
          show two distinct populations of flares, one due to the outbursts, 
          one due to the out-of-outburst emission, which is characterised by 
          fluxes spanning up to 4 orders of magnitude in DR. 
\item By exploiting the higher sensitivity afforded by the \sw/XRT observations and
          by correcting for the sensitivity bias, 
          we find no correlation between the orbital period with any of 
          the duty cycle/activity measurements defined in the soft and hard X--rays. 
          This implies that wide orbits are not characterised by low duty cycles,
          thus answering a long-standing question in SFXT modelling. 
\item The last two findings can be interpreted in terms of mechanisms regulating or 
           inhibiting accretion, such as a propeller effect,  
           magnetic gating, or 
           hot shells of accreted material above the magnetosphere.  
\item The definition of duty cycle is dramatically dependent on the luminosity 
          assumed as lower limit for the calculation.  
\item Our differential count rate distributions indicate that, in order to observe most of 
          the activity of an SFXT, limiting fluxes of at least a decade lower than the sensitivities 
          reached by hard X--ray monitors need to be reached. 
\end{itemize}

Our observations therefore demonstrate that soft X--ray monitoring
campaigns on SFXTs, highly variable sources unpredictably going into outburst,  
can contribute key ingredients, such as dynamical ranges, duty cycles, 
and luminosity distributions, towards characterising them among the general HMXB population.
In particular, given the $\sim 10^{-12}$ erg cm$^{-2}$ s$^{-1}$ sensitivity 
reached in $\sim1$\,ks by the XRT they are uniquely suited to observe most of the 
activity of an SFXT. They also show that 
the most effective way to highlight the SFXT nature of a source is 
the combination of the soft X--ray inactivity duty cycle and dynamical range. 

Finally, we note that, in order to make significant progress towards 
understanding SFXTs as a class and within the HMXB context, 
it is of fundamental importance to continue along this line of investigation, 
by securing long-term soft X--ray data on more SFXTs. 
The fallout of such investigation will be twofold: on one side, 
we shall obtain an increased knowledge on a larger number of individual SFXTs, 
which are on average fainter than the HMXB population and often located 
in crowded, heavily absorbed regions of the sky, and therefore
have not received adequate attention from lower sensitivity soft X--ray 
monitors; on the other side, we can use the combination of 
soft X--ray inactivity duty cycle and dynamical range to select SFXT 
candidates among the HMXB population. 
In this framework, until new insight can be obtained from wide-field, high-sensitivity monitors 
such as those on board LOFT \citep{FerociLOFT_long}, as recently shown by \citet[][]{Bozzo2013:COSPAR_sfxt} and 
\citet[][]{Romano2012:Gamma12_LOFT}, our monitoring campaigns 
are the only viable mean to reach the low luminosities 
($L_{\rm 2-10\,keV}~\sim 10^{33}$--$10^{34}$ erg\,s$^{-1}$) 
required to fully characterise the SFXT phenomenology.

\begin{acknowledgements}
We wholeheartedly thank the {\it Swift} team duty scientists and science planners 
for their courteous efficiency, and A.\ Beardmore, M.\ Capalbi, and H.A.\ Krimm for helpful discussions. 
We also thank our referee, Dr.\ Jan-Uwe Ness, for comments that helped improve the paper.
PR acknowledges contract ASI-INAF I/004/11/0. 
LD thanks Deutsches Zentrum f\"ur Luft und Raumfahrt (Grant FKZ 50 OG 1301). 
\end{acknowledgements}


%

\Online
\onltab{
\setcounter{table}{0} 
\begin{table*} 	
 \tabcolsep 4pt         
 \caption{Observation log for IGR~J08408$-$4503 (J08408).   \label{sfxt10:tab:alldata08408} } 	
 \begin{tabular}{llllr} 
 \hline 
 \hline 
 \noalign{\smallskip} 
Sequence & Instrument/Mode & Start time (UT) &  End time (UT) &  Net Exposure \\
            &     & (yyyy-mm-dd hh:mm:ss) & (yyyy-mm-dd hh:mm:ss) &  (s) \\ 
 \noalign{\smallskip} 
 \hline 
 \noalign{\smallskip} 
00037881013	&XRT/PC &2011-10-20 18:36:10     &2011-10-20 18:52:56     &1003    \\
00037881015	&XRT/PC &2011-10-27 01:18:15     &2011-10-27 01:34:56     &988     \\
00037881016	&XRT/PC &2011-10-30 04:48:18     &2011-10-30 05:03:58     &940     \\
00037881017	&XRT/PC &2011-11-03 16:19:21     &2011-11-03 16:36:57     &988     \\
00037881018	&XRT/PC &2011-11-06 18:08:59     &2011-11-06 18:25:57     &1005    \\
00037881019	&XRT/PC &2011-11-10 23:34:00     &2011-11-10 23:49:58     &938     \\
00037881020	&XRT/PC &2011-11-13 20:38:38     &2011-11-13 20:53:44     &880     \\
00037881021	&XRT/PC &2011-11-17 18:54:18     &2011-11-17 19:09:58     &920     \\
00037881022	&XRT/PC &2011-11-20 19:04:48     &2011-11-20 19:21:56     &1023    \\
00037881023	&XRT/PC &2011-11-24 14:34:00     &2011-11-24 14:35:20     &65      \\
00037881024	&XRT/PC &2011-11-27 16:21:02     &2011-11-27 16:36:58     &935     \\
00037881025	&XRT/PC &2011-12-01 18:34:34     &2011-12-01 18:51:57     &1038    \\
00037881026	&XRT/PC &2011-12-04 04:21:13     &2011-12-04 04:37:56     &993     \\
00037881027	&XRT/PC &2011-12-08 18:46:34     &2011-12-08 19:01:57     &918     \\
00037881028	&XRT/PC &2011-12-11 09:31:44     &2011-12-11 09:48:57     &1025    \\
00037881029	&XRT/PC &2011-12-15 19:25:12     &2011-12-15 19:32:48     &105     \\
00037881030	&XRT/PC &2011-12-18 03:30:41     &2011-12-18 03:45:56     &895     \\
00037881031	&XRT/PC &2011-12-22 13:34:53     &2011-12-22 13:51:56     &1020    \\
00037881032	&XRT/PC &2011-12-25 18:46:59     &2011-12-25 19:03:57     &993     \\
00037881033	&XRT/PC &2011-12-29 02:42:43     &2011-12-29 02:57:51     &311     \\
00037881034	&XRT/PC &2012-01-01 15:41:57     &2012-01-01 15:58:58     &998     \\
00037881035	&XRT/PC &2012-01-05 09:33:53     &2012-01-05 09:48:56     &810     \\
00037881036	&XRT/PC &2012-01-08 12:56:20     &2012-01-08 13:11:58     &928     \\
00037881037	&XRT/PC &2012-01-12 11:57:42     &2012-01-12 12:13:58     &970     \\
00037881038	&XRT/PC &2012-01-15 19:52:01     &2012-01-15 23:28:56     &1020    \\
00037881039	&XRT/PC &2012-01-19 10:33:01     &2012-01-19 10:48:56     &950     \\
00037881040	&XRT/PC &2012-01-22 10:59:58     &2012-01-22 11:13:55     &822     \\
00037881041	&XRT/PC &2012-01-26 20:47:35     &2012-01-26 21:03:55     &968     \\
00037881042	&XRT/PC &2012-01-29 20:59:11     &2012-01-29 21:15:56     &988     \\
00037881043	&XRT/PC &2012-02-02 22:51:42     &2012-02-02 23:07:57     &963     \\
00037881044	&XRT/PC &2012-02-05 21:26:17     &2012-02-05 21:42:57     &983     \\
00037881045	&XRT/PC &2012-02-09 07:14:06     &2012-02-09 07:31:56     &1071    \\
00037881046	&XRT/PC &2012-02-12 07:22:02     &2012-02-12 07:37:57     &938     \\
00037881047	&XRT/PC &2012-02-16 07:43:54     &2012-02-16 08:00:57     &1015    \\
00037881048	&XRT/PC &2012-02-19 17:33:52     &2012-02-19 17:49:57     &945     \\
00037881049	&XRT/PC &2012-02-23 16:06:09     &2012-02-23 16:22:55     &241     \\
00037881050	&XRT/PC &2012-02-26 17:54:06     &2012-02-26 18:12:57     &1106    \\
00037881051	&XRT/PC &2012-03-01 15:06:06     &2012-03-01 15:22:56     &988     \\
00037881052	&XRT/PC &2012-03-04 18:35:29     &2012-03-04 18:52:57     &1036    \\
00037881053	&XRT/PC &2012-03-08 15:37:17     &2012-03-08 15:49:11     &695     \\
00037881054	&XRT/PC &2012-03-11 06:11:13     &2012-03-11 06:23:57     &755     \\
00037881055	&XRT/PC &2012-03-15 06:07:08     &2012-03-15 06:23:56     &993     \\
00037881056	&XRT/PC &2012-03-18 04:36:00     &2012-03-18 04:52:56     &1010    \\
00037881057	&XRT/PC &2012-03-22 06:42:31     &2012-03-22 07:04:57     &1319    \\
00037881058	&XRT/PC &2012-03-25 05:07:24     &2012-03-25 05:23:57     &985     \\
00037881059	&XRT/PC &2012-03-29 19:49:53     &2012-03-29 20:06:56     &1020    \\
00037881060	&XRT/PC &2012-04-01 00:50:59     &2012-04-01 01:00:56     &597     \\
00037881061	&XRT/PC &2012-04-05 09:09:22     &2012-04-05 09:25:57     &983     \\
00037881062	&XRT/PC &2012-04-08 16:02:24     &2012-04-08 16:18:57     &980     \\
00037881063	&XRT/PC &2012-04-12 16:01:18     &2012-04-12 16:18:56     &1058    \\
00037881064	&XRT/PC &2012-04-15 06:40:00     &2012-04-15 06:55:58     &935     \\
00037881065	&XRT/PC &2012-04-19 10:10:24     &2012-04-19 10:26:57     &983     \\
00037881067	&XRT/PC &2012-04-26 02:29:19     &2012-04-26 02:43:57     &875     \\
00037881068	&XRT/PC &2012-04-29 23:17:08     &2012-04-29 23:34:56     &1058    \\
00037881069	&XRT/PC &2012-05-03 06:01:20     &2012-05-03 06:17:58     &973     \\
00037881070	&XRT/PC &2012-05-06 10:55:52     &2012-05-06 11:15:56     &1188    \\
00037881071	&XRT/PC &2012-05-10 22:38:34     &2012-05-10 22:55:57     &1018    \\
00037881072	&XRT/PC &2012-05-13 05:11:24     &2012-05-13 05:27:57     &983     \\
  \noalign{\smallskip}
  \hline
   \end{tabular}
\end{table*} 	
\setcounter{table}{0} 
\begin{table*} 	
 \tabcolsep 4pt         
 \caption{continued.   } 	
 \begin{tabular}{llllr} 
 \hline 
 \hline 
 \noalign{\smallskip} 
Sequence & Instrument/Mode & Start time (UT) &  End time (UT) &  Net Exposure \\
            &     & (yyyy-mm-dd hh:mm:ss) & (yyyy-mm-dd hh:mm:ss) &  (s) \\ 
 \noalign{\smallskip} 
 \hline 
 \noalign{\smallskip} 
 00037881073	&XRT/PC &2012-05-17 15:01:20     &2012-05-17 15:17:58     &988     \\
00037881074	&XRT/PC &2012-05-20 18:25:15     &2012-05-20 18:42:58     &1063    \\
00037881075	&XRT/PC &2012-05-24 13:41:11     &2012-05-24 13:57:57     &995     \\
00037881076	&XRT/PC &2012-05-27 13:56:22     &2012-05-27 14:13:55     &1033    \\
00037881077  &XRT/PC &2012-05-31 14:26:44     &2012-05-31 14:42:57     &953     \\
00037881078  &XRT/PC &2012-06-03 08:13:33     &2012-06-03 08:30:56     &1033    \\
00037881079	&	XRT/PC	&	2012-06-07 20:53:54	&	2012-06-07 21:07:56	&	832	\\
00037881080	&	XRT/PC	&	2012-06-10 03:28:56	&	2012-06-10 03:45:57	&	1018	\\
00037881081	&	XRT/PC	&	2012-06-14 07:08:52	&	2012-06-14 07:25:58	&	1000	\\
00037881082	&	XRT/PC	&	2012-06-17 00:50:04	&	2012-06-17 01:06:57	&	1003	\\
00037881083	&	XRT/PC	&	2012-06-21 02:45:18	&	2012-06-21 03:01:56	&	988	\\
00037881084	&	XRT/PC	&	2012-06-24 02:52:54	&	2012-06-24 03:10:55	&	1066	\\
00037881085	&	XRT/PC	&	2012-06-28 07:37:08	&	2012-06-28 07:52:56	&	930	\\
00037881086	&	XRT/PC	&	2012-07-01 07:48:02	&	2012-07-01 07:50:55	&	163	\\
00037881087	&	XRT/PC	&	2012-07-05 06:20:36	&	2012-07-05 06:36:56	&	965	\\
00037881088	&	XRT/PC	&	2012-07-08 00:25:31	&	2012-07-08 00:41:54	&	980	\\
00037881089	&	XRT/PC	&	2012-07-12 08:26:35	&	2012-07-12 08:44:55	&	1096	\\
00037881091	&	XRT/PC	&	2012-07-19 15:39:17	&	2012-07-19 15:55:55	&	998	\\
00037881092	&	XRT/PC	&	2012-07-22 06:14:22	&	2012-07-22 06:29:55	&	913	\\
00037881093	&	XRT/PC	&	2012-07-26 11:14:45	&	2012-07-26 11:31:56	&	1031	\\
00037881094	&	XRT/PC	&	2012-07-29 07:50:40	&	2012-07-29 08:05:56	&	903	\\
00037881095  &      XRT/PC   &      2012-08-02 21:00:51     &      2012-08-02 21:18:55     &     1078    \\
00037881096  &      XRT/PC   &      2012-08-05 10:03:45     &      2012-08-05 10:20:56     &     1008    \\
  \noalign{\smallskip}
  \hline
   \end{tabular}
\end{table*} 	
}

\onltab{
\setcounter{table}{1} 
\begin{table*} 	
 \tabcolsep 4pt         
 \caption{Observation log for IGR~J16328$-$4726 (J16328).           \label{sfxt10:tab:alldata16328} } 	
 \begin{tabular}{llllr} 
 \hline 
 \hline 
 \noalign{\smallskip} 
Sequence & Instrument/Mode & Start time (UT) &  End time (UT) &  Net Exposure \\
            &     & (yyyy-mm-dd hh:mm:ss) & (yyyy-mm-dd hh:mm:ss) &  (s) \\ 
 \noalign{\smallskip} 
 \hline 
 \noalign{\smallskip} 
00032126001	&XRT/PC   &2011-10-20 04:40:25     &2011-10-20 04:55:58     &913	\\
00032126002	&XRT/PC  &2011-10-24 08:04:20     &2011-10-24 08:19:55     &933	\\
00510701000  &BAT/evt   &2011-12-29 06:35:28 	& 2011-12-29 06:55:30     & 1202     \\  
00032126003	&XRT/PC  &2012-01-16 10:57:12     &2012-01-16 11:13:58     &993	\\
00032126004	&XRT/PC  &2012-01-19 22:29:53     &2012-01-19 23:59:57     &850	\\
00032126005	&XRT/PC  &2012-01-23 08:23:46     &2012-01-23 08:39:34     &920	\\
00032126006	&XRT/PC  &2012-01-26 09:58:36     &2012-01-26 10:13:56     &910	\\
00032126007	&XRT/PC  &2012-01-30 18:14:47     &2012-01-30 18:30:57     &968	\\
00032126008	&XRT/PC  &2012-02-02 18:26:21     &2012-02-02 18:41:56     &915	\\
00032126009	&XRT/PC  &2012-02-06 18:38:59     &2012-02-06 18:55:57     &1010	\\
00032126010	&XRT/PC  &2012-02-09 23:39:26     &2012-02-09 23:54:57     &923	\\
00032126011	&XRT/PC  &2012-02-13 04:51:22     &2012-02-13 05:05:42     &855	\\
00032126012	&XRT/PC  &2012-02-16 20:50:14     &2012-02-16 21:06:56     &995	\\
00032126013	&XRT/PC  &2012-02-20 14:40:38     &2012-02-20 14:55:58     &910	\\
00032126014	&XRT/PC  &2012-02-23 00:30:09     &2012-02-23 02:19:57     &943	\\
00032126015	&XRT/PC  &2012-02-27 02:30:04     &2012-02-27 02:47:57     &1051	\\
00032126016	&XRT/PC  &2012-03-01 13:51:34     &2012-03-01 21:49:55     &1271	\\
00032126017	&XRT/PC  &2012-03-05 18:49:05     &2012-03-05 19:06:56     &1058	\\
00032126018	&XRT/PC  &2012-03-08 14:08:34     &2012-03-08 14:13:02     &258	\\
00032126019	&XRT/PC  &2012-03-12 08:02:31     &2012-03-12 08:18:57     &978	\\
00032126020	&XRT/PC  &2012-03-15 00:28:34     &2012-03-15 00:48:45     &1204	\\
00032126021	&XRT/PC  &2012-03-19 05:37:01     &2012-03-19 05:53:57     &995	\\
00032126022	&XRT/PC  &2012-03-22 21:36:57     &2012-03-22 21:53:58     &998	\\
00032126023	&XRT/PC  &2012-03-26 20:29:03     &2012-03-26 23:44:57     &1103	\\
00032126024	&XRT/PC  &2012-03-29 04:40:21     &2012-03-29 04:56:59     &983	\\
00032126025	&XRT/PC  &2012-04-02 01:29:38     &2012-04-02 01:44:56     &918	\\
00032126026	&XRT/PC  &2012-04-05 14:23:29     &2012-04-05 14:41:58     &1091	\\
00032126027	&XRT/PC  &2012-04-09 17:51:39     &2012-04-09 18:04:56     &777	\\
00032126028	&XRT/PC  &2012-04-12 16:30:58     &2012-04-12 16:48:57     &1078	\\
00032126029	&XRT/PC  &2012-04-16 05:26:59     &2012-04-16 18:47:58     &1103	\\
00032126030	&XRT/PC  &2012-04-19 05:36:36     &2012-04-19 09:11:58     &913	\\
00032126031	&XRT/PC  &2012-04-23 01:23:42     &2012-04-23 01:31:56     &484	\\
00032126032  &XRT/PC  &2012-04-26 07:51:05     &2012-04-26 08:06:58     &940    \\ 
00032126033	&XRT/PC  &2012-04-30 08:07:44     &2012-04-30 08:23:56     &953	\\
00032126034	&XRT/PC  &2012-05-03 05:05:51     &2012-05-03 05:23:57     &1068	\\
00032126035	&XRT/PC  &2012-05-07 02:12:05     &2012-05-07 02:30:56     &1131	\\
00032126036	&XRT/PC  &2012-05-10 02:29:00     &2012-05-10 02:44:58     &958	\\
00032126037	&XRT/PC  &2012-05-14 18:28:13     &2012-05-14 18:43:56     &935	\\
00032126038	&XRT/PC  &2012-05-17 20:14:07     &2012-05-17 20:29:57     &940	\\
00032126040	&XRT/PC  &2012-05-24 11:03:27     &2012-05-24 11:18:57     &915	\\
00032126041	&XRT/PC &2012-05-28 15:59:18	&2012-05-28 16:12:56	&800	\\
00032126042	&XRT/PC &2012-05-31 06:30:50	&2012-05-31 06:46:58	&955	\\
00032126043	&XRT/PC &2012-06-04 06:43:13	&2012-06-04 07:00:56	&1043	\\
00032126044	&	XRT/PC	&	2012-06-07 07:05:48	&	2012-06-07 07:21:56	&	955	\\
00032126045	&	XRT/PC	&	2012-06-11 12:07:27	&	2012-06-11 12:22:57	&	930	\\
00032126046	&	XRT/PC	&	2012-06-14 02:34:48	&	2012-06-14 02:52:56	&	1083	\\
00032126047	&	XRT/PC	&	2012-06-18 10:50:15	&	2012-06-18 11:06:58	&	995	\\
00032126048	&	XRT/PC	&	2012-06-21 20:46:47	&	2012-06-21 21:04:58	&	1083	\\
00032126049	&	XRT/PC	&	2012-06-25 16:15:52	&	2012-06-25 16:31:55	&	953	\\
00032126050	&	XRT/PC	&	2012-06-28 00:12:14	&	2012-06-28 00:18:10	&	344	\\
00042949001	&	XRT/PC	&	2012-06-15 14:11:10	&	2012-06-15 14:20:57	&	582	\\
00042950001	&	XRT/PC	&	2012-06-15 15:47:08	&	2012-06-15 15:56:57	&	574	\\
00032126051	&	XRT/PC	&	2012-07-02 21:20:06	&	2012-07-02 21:36:54	&	1005	\\
00032126053	&	XRT/PC	&	2012-07-09 04:04:07	&	2012-07-09 04:20:55	&	1000	\\
00032126055	&	XRT/PC	&	2012-07-16 10:58:03	&	2012-07-16 11:15:54	&	1061	\\
00032126056	&	XRT/PC	&	2012-07-19 14:21:55	&	2012-07-19 14:39:56	&	1073	\\
00032126057	&	XRT/PC	&	2012-07-23 11:28:27	&	2012-07-23 13:18:54	&	1204	\\
00032126058	&	XRT/PC	&	2012-07-26 04:55:06	&	2012-07-26 05:08:53	&	805	\\
  \noalign{\smallskip}
 \hline
 \noalign{\smallskip}
  \end{tabular}
  \end{table*}
\setcounter{table}{1} 
\begin{table*} 	
 \tabcolsep 4pt         
 \caption{continued. } 	
 \begin{tabular}{llllr} 
 \hline 
 \hline 
 \noalign{\smallskip} 
Sequence & Instrument/Mode & Start time (UT) &  End time (UT) &  Net Exposure \\
            &     & (yyyy-mm-dd hh:mm:ss) & (yyyy-mm-dd hh:mm:ss) &  (s) \\ 
 \noalign{\smallskip} 
 \hline 
 \noalign{\smallskip} 
00032126059	&	XRT/PC	&	2012-07-30 00:14:53	&	2012-07-30 00:29:56	&	898	\\
 00032126060  &      XRT/PC   &      2012-08-02 13:19:17     &     2012-08-02 13:35:55     &      983     \\
 00032126061	&	XRT/PC &	2012-08-06 23:02:02	&	2012-08-06 23:19:55	&	1058	\\
00032126062	&	XRT/PC &	2012-08-09 21:48:59	&	2012-08-09 22:06:55	&	1066	\\
00032126063  &     XRT/PC &       2012-08-13 23:42:50     &       2012-08-13 23:59:55     &       1013    \\
00032126064  &     XRT/PC &       2012-08-16 02:52:13     &       2012-08-16 03:08:54     &       978     \\
00032126065  &     XRT/PC &       2012-08-20 11:11:08     &       2012-08-20 11:29:54     &       1106    \\
00032126066  &     XRT/PC &      2012-08-23 16:12:42     &       2012-08-23 16:33:54     &       1256    \\
00032126067  &     XRT/PC &       2012-08-27 06:48:58     &       2012-08-27 07:03:55     &       893     \\
00032126068  &     XRT/PC &       2012-08-30 05:20:38     &       2012-08-30 05:34:56     &       837     \\
00032126069     &       XRT/PC  &       2012-09-03 00:45:13     &       2012-09-03 01:00:56     &       928     \\
00032126070     &       XRT/PC  &       2012-09-06 02:27:30     &       2012-09-06 02:42:55     &       918     \\
00032126072     &       XRT/PC  &       2012-09-13 07:34:36     &       2012-09-13 07:49:54     &       913     \\
00032126073     &       XRT/PC  &       2012-09-17 01:21:14     &       2012-09-17 01:34:54     &       802     \\
00032126074     &       XRT/PC  &       2012-09-20 04:39:22     &       2012-09-20 06:24:55     &       1008    \\
00032126075     &       XRT/PC  &       2012-09-24 01:39:44     &       2012-09-24 01:54:54     &       903     \\
00032126076     &       XRT/PC  &       2012-09-27 02:14:04     &       2012-09-27 02:26:56     &       760     \\
00032126077     &       XRT/PC  &       2012-10-01 00:36:54     &       2012-10-01 00:50:54     &       820     \\
00032126078     &       XRT/PC  &       2012-10-04 10:14:26     &       2012-10-04 10:30:54     &       965     \\
00032126079     &       XRT/PC  &       2012-10-08 00:56:38     &       2012-10-08 01:08:55     &       730     \\
00032126081     &       XRT/PC  &       2012-10-15 14:10:22     &       2012-10-15 14:26:55     &       978     \\
00032126082     &       XRT/PC  &       2012-10-18 22:24:46     &       2012-10-18 22:40:54     &       955     \\
00032126083     &       XRT/PC  &       2012-10-22 21:04:31     &       2012-10-22 21:09:54     &       308     \\
00032126084     &       XRT/PC  &       2013-09-02 02:39:41     &       2013-09-02 02:56:54     &       1025    \\
00032126085     &       XRT/PC  &       2013-09-05 13:55:41     &       2013-09-05 14:13:54     &       1083    \\
00032126086     &       XRT/PC  &       2013-09-09 06:22:31     &       2013-09-09 06:38:57     &       298     \\
00032126087     &       XRT/PC  &       2013-09-12 12:31:55     &       2013-09-12 12:47:48     &       940     \\
00032126088     &       XRT/PC  &       2013-09-16 12:45:00     &       2013-09-16 12:59:55     &       893     \\
00032126089     &       XRT/PC  &       2013-09-19 00:03:17     &       2013-09-19 00:17:56     &       815     \\
00032126090     &       XRT/PC  &       2013-09-23 06:19:13     &       2013-09-23 06:34:53     &       920     \\
00032126091     &       XRT/PC  &       2013-09-26 20:47:21     &       2013-09-26 21:01:54     &       855     \\
00032126092     &       XRT/PC  &       2013-09-30 09:58:40     &       2013-09-30 10:16:55     &       1096    \\
00032126093     &       XRT/PC  &       2013-10-03 01:57:38     &       2013-10-03 02:14:54     &       860     \\
00032126094     &       XRT/PC  &       2013-10-07 14:51:09     &       2013-10-07 15:06:54     &       938     \\
00032126095     &       XRT/PC  &       2013-10-10 03:59:21     &       2013-10-10 04:11:53     &       737     \\
00032126096     &       XRT/PC  &       2013-10-14 21:35:58     &       2013-10-14 21:49:55     &       815     \\
00032126097     &       XRT/PC  &       2013-10-17 03:57:58     &       2013-10-17 04:12:56     &       875     \\
00032126098     &       XRT/PC  &       2013-10-21 05:42:54     &       2013-10-21 06:04:56     &       1319    \\
00032126099     &       XRT/PC  &       2013-10-24 05:41:12     &       2013-10-24 05:57:55     &       1003    \\
   \noalign{\smallskip}
  \hline
   \end{tabular}
\end{table*} 	
}

\onltab{
\setcounter{table}{2} 
\begin{table*} 	
 \tabcolsep 4pt         
 \caption{Observation log for  IGR~J16465$-$4507 (J16465). 
          \label{sfxt10:tab:alldata16465} }	
 \begin{tabular}{llllr} 
 \hline 
 \hline 
 \noalign{\smallskip} 
Sequence & Instrument/Mode & Start time (UT) &  End time (UT) &  Net Exposure \\
            &     & (yyyy-mm-dd hh:mm:ss) & (yyyy-mm-dd hh:mm:ss) &  (s) \\ 
 \noalign{\smallskip} 
 \hline 
 \noalign{\smallskip} 
00032617001	&	XRT/PC	&	2013-01-20 04:09:45	&	2013-01-20 04:25:56	&	935	\\
00032617002	&	XRT/PC	&	2013-01-23 04:20:53	&	2013-01-23 04:36:55	&	938	\\
00032617003	&	XRT/PC	&	2013-01-27 01:20:53	&	2013-01-27 01:35:55	&	895	\\
00032617004	&	XRT/PC	&	2013-01-30 22:34:44	&	2013-01-30 22:50:45	&	948	\\
00032617005	&	XRT/PC	&	2013-02-03 19:12:14	&	2013-02-03 19:29:54	&	1043	\\
00032617006	&	XRT/PC	&	2013-02-06 17:57:37	&	2013-02-06 18:15:55	&	1073	\\
00032617007	&	XRT/PC	&	2013-02-10 19:48:15	&	2013-02-10 20:06:53	&	1098	\\
00032617008	&	XRT/PC	&	2013-02-13 00:31:56	&	2013-02-13 00:47:53	&	938	\\
00032617009	&	XRT/PC	&	2013-02-17 13:22:56	&	2013-02-17 13:38:53	&	953	\\
00032617010	&	XRT/PC	&	2013-02-20 21:39:44	&	2013-02-20 21:57:55	&	1083	\\
00032617011	&	XRT/PC	&	2013-02-24 11:58:31	&	2013-02-24 12:14:54	&	978	\\
00032617012	&	XRT/PC	&	2013-02-27 23:19:43	&	2013-02-27 23:35:54	&	950	\\
00032617013	&	XRT/PC	&	2013-03-03 20:14:01	&	2013-03-03 20:29:54	&	943	\\
00032617014	&	XRT/PC	&	2013-03-06 20:31:45	&	2013-03-06 20:47:55	&	955	\\
00032617015	&	XRT/PC	&	2013-03-10 09:35:53	&	2013-03-10 09:52:56	&	1013	\\
00032617016	&	XRT/PC	&	2013-03-13 00:03:17	&	2013-03-13 00:17:56	&	820	\\
00032617017	&	XRT/PC	&	2013-03-17 00:25:24	&	2013-03-17 00:41:47	&	983	\\
00032617018	&	XRT/PC	&	2013-03-20 22:42:49	&	2013-03-20 22:59:55	&	1008	\\
00032617019	&	XRT/PC	&	2013-03-24 13:23:47	&	2013-03-24 13:39:55	&	958	\\
00032617020	&	XRT/PC	&	2013-03-27 02:24:54	&	2013-03-27 02:41:54	&	1010	\\
00032617021	&	XRT/PC	&	2013-03-31 11:52:15	&	2013-03-31 12:08:55	&	993	\\
00032617022	&	XRT/PC	&	2013-04-03 02:37:49	&	2013-04-03 02:53:54	&	958	\\
00032617023	&	XRT/PC	&	2013-04-07 18:38:59	&	2013-04-07 18:54:55	&	943	\\
00032617024	&	XRT/PC	&	2013-04-10 15:25:32	&	2013-04-10 15:41:55	&	960	\\
00032617026	&	XRT/PC	&	2013-04-17 01:32:52	&	2013-04-17 01:48:55	&	955	\\
00032617028	&	XRT/PC	&	2013-04-24 19:08:05	&	2013-04-24 19:28:54	&	1249	\\
00032617029	&	XRT/PC	&	2013-04-28 09:53:42	&	2013-04-28 10:12:55	&	1153	\\
00032617030	&	XRT/PC	&	2013-05-01 14:44:28	&	2013-05-01 14:59:56	&	908	\\
00032617031	&	XRT/PC	&	2013-05-05 00:32:17	&	2013-05-05 00:53:54	&	1291	\\
00032617032	&	XRT/PC	&	2013-05-08 05:20:35	&	2013-05-08 05:36:55	&	960	\\
00032617033	&	XRT/PC	&	2013-05-12 21:55:42	&	2013-05-12 23:41:56	&	945	\\
00032617034	&	XRT/PC	&	2013-05-15 21:36:49	&	2013-05-15 21:42:53	&	364	\\
00032617035	&	XRT/PC	&	2013-05-19 01:12:11	&	2013-05-19 01:18:05	&	351	\\
00032617036	&	XRT/PC	&	2013-05-22 09:07:41	&	2013-05-22 09:23:54	&	953	\\
00032617037	&	XRT/PC	&	2013-05-26 09:16:54	&	2013-05-26 09:33:54	&	1010	\\
00032617038	&	XRT/PC	&	2013-05-29 09:22:51	&	2013-05-29 09:40:55	&	1063	\\
00032617039	&	XRT/PC	&	2013-06-02 09:31:33	&	2013-06-02 09:47:53	&	978	\\
00032617040	&	XRT/PC	&	2013-06-05 19:22:49	&	2013-06-05 19:38:55	&	958	\\
00032617041	&	XRT/PC	&	2013-06-09 09:56:31	&	2013-06-09 10:02:55	&	379	\\
00032617042	&	XRT/PC	&	2013-06-12 16:21:26	&	2013-06-12 16:36:54	&	920	\\
00032617043	&	XRT/PC	&	2013-06-16 03:38:35	&	2013-06-16 03:54:56	&	958	\\
00032617045	&	XRT/PC	&	2013-06-23 02:03:29	&	2013-06-23 02:21:55	&	1096	\\
00032617046	&	XRT/PC	&	2013-06-26 20:03:08	&	2013-06-26 20:18:54	&	943	\\
00032617047	&	XRT/PC	&	2013-06-30 00:48:38	&	2013-06-30 01:04:53	&	958	\\
00032617048	&	XRT/PC	&	2013-07-03 23:15:52	&	2013-07-03 23:33:55	&	1063	\\
00032617049	&	XRT/PC	&	2013-07-07 18:33:51	&	2013-07-07 18:50:54	&	1000	\\
00032617050	&	XRT/PC	&	2013-07-10 01:06:00	&	2013-07-10 06:02:56	&	870	\\
00032617051	&	XRT/PC	&	2013-07-14 06:14:33	&	2013-07-14 11:19:54	&	903	\\
00032617052	&	XRT/PC	&	2013-07-17 19:11:31	&	2013-07-17 19:28:54	&	1038	\\
00032617053	&	XRT/PC	&	2013-07-21 14:13:37	&	2013-07-21 14:29:55	&	973	\\
00032617054	&	XRT/PC	&	2013-07-24 06:24:07	&	2013-07-24 06:40:55	&	983	\\
00032617056	&	XRT/PC	&	2013-07-31 11:30:13	&	2013-07-31 11:47:53	&	1053	\\
00032617057	&	XRT/PC	&	2013-08-04 21:16:14	&	2013-08-04 22:36:54	&	1076	\\
00032617058	&	XRT/PC	&	2013-08-07 06:51:42	&	2013-08-07 07:09:55	&	1091	\\
00032617059	&	XRT/PC	&	2013-08-11 02:10:07	&	2013-08-11 02:26:55	&	988	\\
00032617060	&	XRT/PC	&	2013-08-14 21:16:47	&	2013-08-14 21:31:55	&	885	\\
00032617061  &       XRT/PC  &       2013-08-18 15:03:58     &       2013-08-18 15:21:54     &       1058    \\
00032617062  &       XRT/PC  &       2013-08-21 03:58:44     &       2013-08-21 04:15:55     &       1020    \\
00032617063  &       XRT/PC  &       2013-08-25 04:05:54     &       2013-08-25 05:52:54     &       890     \\
00032617064  &       XRT/PC  &       2013-08-28 12:22:54     &       2013-08-28 18:51:56     &       820     \\
00032617065  &       XRT/PC  &       2013-09-01 17:03:18     &       2013-09-01 17:20:56     &       1046    \\
  \noalign{\smallskip}
  \hline
   \end{tabular}
\end{table*} 	
}

\onltab{
\setcounter{table}{3} 
\begin{table*}  
 \tabcolsep 4pt         
 \caption{Observation log for  outburst data. 
          \label{sfxt10:tab:alldata_outburst} } 
 \begin{tabular}{llllllr}
 \hline 
 \hline 
 \noalign{\smallskip} 
Source  &Nickname& Sequence & Mode & Start time (UT) &  End time (UT) &  Net Exposure \\
             &&          &     & (yyyy-mm-dd hh:mm:ss) & (yyyy-mm-dd hh:mm:ss) &  (s) \\ 
 \noalign{\smallskip} 
 \hline 
 \noalign{\smallskip} 
IGR~J08408$-$4503 & J08408
          & 00559642000     &       XRT/WT  &      2013-07-02 08:13:08     &       2013-07-02 11:30:45     &       478     \\
          && 00559642000     &       XRT/PC  &       2013-07-02 08:20:08     &       2013-07-02 11:49:07     &       4298    \\
          && 00037881097     &       XRT/PC  &       2013-07-03 11:40:45     &       2013-07-03 11:57:20     &       975     \\
          && 00037881098     &       XRT/PC  &       2013-07-04 06:35:35     &       2013-07-04 06:52:11     &       978     \\
          && 00037881100     &       XRT/PC  &       2013-07-05 01:56:49     &       2013-07-05 02:14:55     &       1071    \\
          && 00037881101     &       XRT/PC  &       2013-07-06 17:56:49     &       2013-07-06 19:39:55     &       1043    \\
          && 00037881102     &       XRT/PC  &       2013-07-07 00:25:08     &       2013-07-07 00:41:56     &       995     \\
 \noalign{\smallskip}
IGR~J16328$-$4726   &J16328
           & 00354542000     &XRT/PC     &2009-06-10 08:01:15     & 2009-06-10 13:00:51     & 8181      \\
          && 00354542001     &XRT/PC      &2009-06-11 15:47:45     & 2009-06-11 16:04:18     & 993       \\
          && 00354542002     &XRT/PC     &2009-06-12 15:50:48     & 2009-06-12 16:07:21     & 985       \\
          && 00354542003     &XRT/PC     &2009-06-13 15:56:57     & 2009-06-13 16:13:30     & 975      \\  
          && 00354542004     &XRT/PC      &2009-06-14 00:13:26     & 2009-06-15 21:11:56     & 2827     \\  
 \noalign{\smallskip}
AX~J1841.0$-$0536  & J1841
          & 00524364000     &       XRT/WT  &       2012-06-14 19:18:55     &       2012-06-14 20:22:09     &       257     \\
          && 00524364000     &       XRT/PC  &       2012-06-14 20:22:10     &       2012-06-14 20:57:44     &       2111    \\
          && 00030988115     &       XRT/PC  &       2012-06-15 10:40:44     &       2012-06-15 12:52:45     &       700     \\
          && 00030988117     &       XRT/PC  &       2012-06-17 11:17:45     &       2012-06-17 12:58:56     &       1018    \\
          && 00030988118     &       XRT/PC  &       2012-06-18 15:57:33     &       2012-06-18 16:06:57     &       554     \\
          && 00030988119     &       XRT/PC  &       2012-06-16 16:03:51     &       2012-06-16 16:15:58     &       705     \\
          && 00030988120     &       XRT/PC  &       2012-06-19 03:32:33     &       2012-06-19 19:36:56     &       1246    \\ 
\noalign{\smallskip}
\hline
\noalign{\smallskip}
 \end{tabular}
  \end{table*} 
}

\end{document}